\documentclass[showpacs,preprint,amssymb]{revtex4}
\usepackage{graphicx}
\usepackage{amsmath}
\usepackage{dcolumn}
\usepackage{bm}
\newcommand\dd{\ensuremath{\text{d}}}
\def\bpm{\begin{pmatrix}}
\def\epm{\end{pmatrix}}
\def\be{\begin{equation}} 
\def\ee{\end{equation}}
\def\bea{\begin{eqnarray}} 
\def\eea{\end{eqnarray}}
\def\line{\hbox to \hsize}    
\def\frac #1#2{{#1\over #2}}

\def \e{{\bf e}}    
\def \a{{\bf a}}
\def \b{{\bf b}}
\def \c{{\bf c}}

\def \n{{\bf n}}

\def\bz{{\bar z}}

\def \ket #1{{\vert #1\rangle}}
\def \bra #1{{\langle #1\vert}}
\def \brak #1#2{{\langle#1\vert#2\rangle}}
\def\eval #1#2#3{{\langle#1\vert#2\vert#3\rangle}} 
\def\vev #1{{\langle #1\rangle}}
\def\1{\mbox{\bf 1}}
\def\bm#1{\mbox{\boldmath$#1$}} 
\def\levelonelist{
        \begin{list}{\mybulA}%
                        {
        \setlength{\topsep}{0pt}
        \setlength{\parsep}{0pt}
        \setlength{\partopsep}{0pt}
        \setlength{\itemsep}{0pt}
                        }
                }
\def\leveltwolist{
        \begin{list}{\mybulB}%
                        {
        \setlength{\topsep}{0pt}
        \setlength{\parsep}{0pt}
        \setlength{\partopsep}{0pt}
        \setlength{\itemsep}{0pt}
                        }
                }

\def\el{\end{list}}

\begin{document}

\title{ Non-Abelian Berry transport, spin coherent states, and Majorana points.}

 \author{ YUN LIU}

\affiliation{University of Illinois, Department of Physics\\ 1110 W. Green St.\\
Urbana, IL 61801 USA\\E-mail: yunliu1@illinois.edu}

\author{ABHISHEK ROY}
\affiliation{
University of Illinois, Department of Physics\\ 1110 W. Green St.\\
Urbana, IL 61801 USA
\\E-mail: aroy2@illinois.edu}

 \author{ MICHAEL STONE}

\affiliation{University of Illinois, Department of Physics\\ 1110 W. Green St.\\
Urbana, IL 61801 USA\\E-mail:  m-stone5@illinois.edu}   

\begin{abstract}  
 We consider the adiabatic evolution of Kramers degenerate pairs of  spin states in a half-integer spin   molecular magnet as the molecule  is slowly   rotated.  To reveal the full details  the of the quantum evolution, we use  Majorana's parametrisation of a general state in the  $2j+1$ dimensional Hilbert space in terms of $2j$ Majorana points.   We show that the intricate  motion of the Majorana points may be  described by  a classical hamiltonian which is of the same form, but of quite different origin,  as that which appears in the  spin-coherent-state path integral.  
 As an illustration we consider  molecular magnets of the $j=9/2$  Mn4 family and compute the frequency  with which the magnetization varies. This frequency is generally  different from the frequency  of the rotation.  
 
 \end{abstract}  


\maketitle

\section{Introduction}
\label{SEC:introduction}

The controlled and measurable manipulation of quantum spin states  is the basis of NMR with its abundance of practical applications.   Another system   in which  spins may be manipulated and measured  is provided by molecular magnets \cite{wernsdorfer-nature}. These materials  consist of  large organic molecules  containing clusters of transition elements such as Mn or Fe, whose unpaired electron spins lock together to form a single large ($j\sim 10$) spin.  The best known examples  are Mn12, Fe8, and Mn4,  for which $j=10$, $10$ and $9/2$ respectively.  (The nomenclature focusses on the  magnetic atoms.   Mn12, for example,  is shorthand for the polyacetate  \hbox{Mn$_{12}$O$_{12}$(CH$_3$COO)$_{16}\cdot$(H$_2$O)$_4$}.)  The organic ligands surrounding the magnetic cluster  serve to reduce the interaction between clusters  so that each molecular spin behaves almost  independently --- yet,  when the molecules have a common orientation and all experience the same external forces,   the total magnetic moment is the sum of many identical molecular  moments and so the  spin direction may  be   observed. 
The energy of the  spin ${\bf J}$ depends on its orientation with respect to the host molecule, and one may  think of the tip of the spin arrow as   moving on a sphere through a  potential landscape possessing  hills, valleys and saddlepoints. 
This  potential landscape can  be   manipulated  by applying an external magnetic field. In Fe8,  for example,  the rate of quantum spin tunneling  between  eigenstates  located in separate potential  wells has been measured  {\it via} the Landau-Zener  crossing   of  the levels as the external field is varied \cite{wernsdorfer99}. 

An alternative  route to manipulating the spin is simply to  rotate the system. In most cases the magnetization  will  merely  co-rotate with the molecule, but when the spin ground-state is degenerate we have   more interesting possibilities.  If the rotation is slow compared to the separation of the degenerate ground state from  the first  excited state, the evolution will be   described by  non-abelian Berry transport 
\cite{wilczek-zee,moody-shapere-wilczek}.   Now a  doubly-degenerate ground state is guaranteed by Kramers' theorem whenever the total spin is a half integer, and the Hamiltonian $H({\bf J})$ governing the large spin's interaction with the host molecule  is invariant under time reversal  \cite{mead}.       
The aim of this  paper is to  explore how  these  Kramers-degerate pairs   evolve as a general anisotropic molecule is rotated.

When dealing with the $2j+1$ dimensional Hilbert space of a large spin, the expectation   $\eval{\psi}{{\bf J}}{\psi}$ reveals only a small part of the quantum information encoded in   the spin state $\ket{\psi}$.   A useful tool \cite{leboeuf} for visualizing the complete quantum state  is provided by the location of the zeros on the unit sphere of the spin-coherent-state wavefunction $\psi(z)$. These zeros are the antipodes of  ``Majorana points''  that can be thought of the directions of a set of $2j$  spin-$1/2$'s  that compose the spin-$j$.

In section \ref{SEC:majorana} we will review   Majorana's parametrisation \cite{majorana} of  a general spin-$j$ state  and explain its connection with spin-coherent state wavefunctions. In section \ref{SEC:berry} we will derive expressions for the non-abelian Berry connection one-form in terms of the Majorana parametrisation, and show how they simplify when we restrict to Kramers degenerate pairs of states. In sections \ref{SEC:kato}  and \ref{SEC:action} we introduce an alternative way of thinking about the adiabatic Berry transport for Kramers pairs, and  how the evolution of the state  is described by a classical hamiltonian dynamical system.  In the last section \ref{SEC:illustration}  we show how even  simple anisotropies can give rise to  intricate  state evolutions  as a molecule is rotated, and that this intricacy may have experimentally observable  consequences. In particular we will see   that a steady period $T$ rotation of the molecule  can  give rise to motions of the spin direction  that are again periodic, but with a period that typically  differs from $T$. 

\section{Majorana parametrisation and  spin coherent states}
\label{SEC:majorana}

It is a familiar fact that when a collection of $2j$ spin-1/2 particles  are combined into a single quantum system  the spin-$j$ representation of ${\rm SU}(2)$ resides in the space of totally symmetric tensors. Elements of a symmetric tensor space ${\rm Sym} \left[V^{\otimes r}\right]$ are sums of the form 
\be
A^{i_1 i_2\ldots i_{r}} \e_{i_1}\odot \e_{i_2}\odot\cdots \odot \e_{i_{r}},
\ee
where the $\e_i$ are the basis vectors for $V$, and the  symmetric tensor product ``$\odot$'' is defined to commutative and distributive:
\be
\a\odot \b=\b\odot \a, \quad \a\odot(\lambda \b+\mu\c)= \lambda \,(\a\odot \b)+\mu\,( \a\odot \c).
\ee

In \cite{majorana} Majorana showed    that an  arbitrary  spin-$j$ state ${\bm \Psi}\in {\rm Sym}[V_{1/2}^{\otimes 2j}]$ can be decomposed as 
\be
{\bm \Psi}= {\bm \psi}_{i_1}\odot \cdots \odot {\bm \psi}_{i_{2j}},
\ee
where the \be
{\bm \psi}_{i}= \alpha_i\,\e_1+ \beta_i\,  \e_2
\ee
are  a set of  $2j$ spin-1/2 spinors with ${\bf e}_1\equiv\ket{\!\!\uparrow}$, ${\bf e}_2=\ket{\!\!\downarrow}$. This result is perhaps surprising. Most elements of a general tensor product space (whether symmetric or not) cannot be  decomposed into a single product; they  can only be expressed as  a {\it sums\/}  of  products.  Majorana's result was rediscovered by Schwinger \cite{schwinger}  and by   Rabi \cite{rabi},  and later reviewed by Bloch and Rabi \cite{bloch-rabi}. Schwinger's paper is historically  interesting as in his equation (9)  he draws attention to a previously omitted term in the equations describing the evolution of a spin in  a time-varying field. This term is now recognizable as the  ``Berry Phase.''  Majorana's original paper  is similarly   notable in that it contains an independent  derivation of the non-adiabatic  level-crossing probability  that is traditionally  attributed to Landau, Stuckleberg and Zener.

Majorana's recipe for his  decomposition is  as follows:  Given a spin-$j$ state 
\be
{\bm\Psi}\equiv \ket{ \Psi}= \sum_m \ket {j,m}\brak{j,m}{\Psi}
\ee
 we   construct    a set of $2j+1$  coefficients 
\be
 c_{j-m} = \brak{j,m}{\Psi}  \sqrt{\frac{(2j)!}{(j+m)!(j-m)!} } , \quad m=-j, -j+1,\ldots ,j-1,j
 \label{EQ:coefficients}
 \ee
and 
the degree $2j$ polynomial equation 
\be
P_\Psi(z)=\sum_{n=0}^{2j} z^{2j-n} c_{n}=0.
\ee
If  the $2j$  roots  of this polynomial are denoted by $z_i$, then the $z_i$  are related to the  coefficients of the spin-$1/2$ factors by 
\be
z_i = -\beta_i/\alpha_i.
\ee
We can extract   the individual coefficients  from their ratios by normalizing so that \hbox{$|\alpha_i|^2+|\beta_i|^2=1$}, but a   choice of phase remains for each ${\bm \psi}_i$. These  phases can be swapped amongst the factors leaving only an overall phase to be matched with that of ${\bm \Psi}$. 

 It may happen that one or more of the leading  coefficients $c_{n}$ are zero. The polynomial equation then has degree less than $2j$ and possesses  fewer than $2j$ roots. 
In this case  it is  convenient to  extend the complex plane  to the Riemann sphere, and regard the missing  roots  as being   at infinity. Their  corresponding $\alpha_i$ are then zero.   

To understand why the recipe works, we   use the commuting property of the symmetric tensor factors to  allow the shorthand 
\be
\underbrace{\e_1\odot\cdots\odot \e_1}_{j+m\,\,{\rm factors}}  \odot\underbrace{\e_2\odot\cdots\odot \e_2}_{j-m\, \,{\rm factors}}= \e_1^{j+m} \e_2^{j-m}.
\ee
To connect an expansion in terms of the $\e_1^{j+m} \e_2^{j-m}$ with  the more conventional expansion
\be
{\bm \Psi}\equiv  \ket{\Psi}= \sum_m\ket{j,m}\brak{j,m}{\Psi}
\ee
we need to define an inner product on the symmetric tensor-product space. The one that is induced naturally  from the spin-1/2 product sets 
\be
\brak{  \e_1^{j+m} \e_2^{j-m}}{\e_1^{j+m'} \e_2^{j-m'}}= \delta_{mm'} \frac{(j+m)!(j-m)!}{(2j)!}.
\ee
From this  we identify
\be
\ket{j,m} = \sqrt{\frac{(2j)!}{(j+m)!(j-m)!} }\e_1^{j+m} \e_2^{j-m}.
\ee
This natural inner product arises from regarding  the symmetric tensor products as being symmetrized  conventional tensor products. For example,
\be
\e_1\odot\e_1\odot \e_2\equiv  \frac 13 \left(\e_1\otimes \e_1\otimes \e_2+ \e_1\otimes \e_2\otimes \e_1+\e_2\otimes \e_1\otimes \e_1\right).
\ee
In the general case, the 3 is replaced by
\be
N_{jm}=  \frac{(2j)!}{(j+m)!(j-m)!},
\ee
which is the number of distinct terms generated when we symmetrize
\be
 \e_{i_1}\odot \e_{i_2}\odot\cdots \odot \e_{i_{2j}}= \frac{1}{(2j)!}\sum_{\pi \in {\mathfrak S}_{2j}}   \e_{i_{\pi(1)}}\otimes \e_{i_{\pi(2)}}\otimes\cdots \otimes \e_{i_{\pi(2j)}}
 \ee
 by summing over the $(2j)!$ elements of the permutation group ${\mathfrak S}_{2j}$.
 The $N_{jm}$ distinct terms are mutually orthonormal with regard to the standard  inner product on the  unsymmetrized $[V_{1/2}]^{\otimes 2j}$, so computing the inner  product from this conventional tensor-product side  makes   
\be
\brak{  \e_1^{j+m} \e_2^{j-m}}{\e_1^{j+m} \e_2^{j-m}}=
(N_{jm})^{-2}  N_{jm}= (N_{jm})^{-1}.
\ee
To confirm the identification,  observe  that  $J_- \e_1=\e_2$, $ J_- \e_2=0$, and so
\bea
J_- \ket{j,m} &=&  \sqrt{\frac{(2j)!}{(j+m)!(j-m)!} } J_- (\e_1^{j+m} \e_2^{j-m})\nonumber\\
&=& \sqrt{\frac{(2j)!}{(j+m)!(j-m)!} } (j+m)  \e_1^{j+m-1} \e_2^{j-m+1}\nonumber\\
&=&\sqrt{\frac{(2j)!}{(j+m)!(j-m)!} }   (j+m) \sqrt{\frac{(j+m-1)!(j-m+1)!}{(2j)!} }\ket{j,m-1}\nonumber\\
&=& \sqrt{ (j+m)(j-m+1)} \ket{j,m-1}
\eea 
as it should.

Now let ${\bm \psi}_i = \alpha_i \e_1+\beta_i \e_2$. Set $z_i= -\beta_i/\alpha_i$, and  expand out a decomposable state as
\be
{\bm \Psi}\equiv {\bm \psi}_{i_1}\odot \cdots \odot {\bm \psi}_{i_{2j}}=   \left(\prod_{i=1}^{2j} \alpha_i\right) a_{j-m}\e_1^{j+m} \e_2^{j-m},
\ee
where $a_{j-m}$ is the $(j-m)$-th elementary symmetric function of the $-z_i$. In other words 
\be
\sum_{n=0}^{2j} z^{2j-n} a_{n}=\prod_{i=1}^{2j} (z-z_i).  
\ee
We therefore have 
\be
\brak{j,m}{\Psi} =  \left(\prod_{i=1}^{2j} \alpha_i\right) a_{j-m} \sqrt{ \frac{(j+m)!(j-m)!}{(2j)!}}.
\ee
or 
\be
 a_{j-m} = \brak{j,m}{\Psi}  \sqrt{\frac{(2j)!}{(j+m)!(j-m)!} }  \left(\prod_{i=1}^{2j} \alpha_i\right)^{-1}.
 \ee
The common factor   $\prod_{i=1}^{2j} \alpha_i= \brak{j,j}{\Psi}$  does not affect the location of the zeros, so we can replace the $a_{j-m}$ by Majorana's  $c_{j-m}$ (which are $\brak{j,j}{\Psi}a_{j-m}$). Thus  given the components $\brak{j,m}{\Psi}$ we can solve  for the roots of $P_\Psi(z)=\sum_{n=0}^{2j} z^{2j-n} c_{n}=0$ and   construct the $\alpha_i$, $\beta_i$ up to  an overall  phase as before.

The Majorana  polynomial $P_\Psi(z)$ for a given $\ket{\Psi}$  has a physical interpretation as the holomorphic  spin-coherent-state wavefunction representation of  $\ket{\Psi}$.   
This wavefunction is obtained by first  defining  a family of spin coherent states
\be
\ket{z} = \exp(\bz  J_+) \ket{j,-j},\quad  \bra{z} =\bra{j,-j}\exp(z J_-)\equiv \ket{z}^\dagger.
\ee
These  states  are  not normalized, but  have the advantage  that the $\bra{z}$ are holomorphic in the parameter $z$ --- {\it i.e.\/}, they depend on $z$ but not on $\bz$.  The coherent-state wavefunction  $\psi(z) \stackrel{\rm def}{=} \brak{z}{\psi}$ corresponding to a state $\ket{\psi}$ is then a holomorphic function.
The   label  $z$ can  be identified as  the  complex sterographic co-ordinate of the spin direction on the two-sphere, with  $z=0$ corresponding  to the south pole (spin down)  and $z=\infty$ to the north pole (spin up). (Strictly speaking $\ket{\infty}$ is undefined, and one needs two families of coherent states with a transition function to capture every spin direction --- see for example \cite{stone-park-garg}. This technicality is not relevant for what follows however.)

The inner product of two  states $\ket{\psi}$  and $\ket{\chi}$ may be evaluated in terms of their  wavefunctions as   
\be
\brak{\psi}{\chi}  = 
\frac {2j+1}{2\pi i } \int_{\mathbb C}  \frac{d\bz \wedge dz }{(1+| z|^2)^{2j+2}}\,\overline {\psi(z)}\chi(z).
\label{EQ:spin_inner_product}
\ee
Normalizable wavefunctions are therefore  polynomials in $z$ of degree less than or equal to $2j$. In particular 
\be
\brak{z}{j,m}=\sqrt{\frac{(2j)!}{(j-m)!(j+m)!}}\, z^{j+m},
\label{EQ:z_basis}
\ee
and the  wavefunction of  coherent state $\ket{\zeta}$ is the polynomial 
\be
\brak{z}{\zeta} = (1+ z \bar \zeta)^{2j}.
\ee

When acting on the holomorphic wavefunctions then   $\mathfrak{su}(2)$ generators  $J_3$ and $J_{\pm}=J_1\pm iJ_2$  become
\bea
J_+ &\to& -z^2\frac{\partial}{\partial z} +2jz\nonumber\\
 J_-& \to &\phantom z \frac{ \partial}{\partial z},\nonumber \\
 J_3&\to & z\frac{\partial}{\partial z} -j.
 \eea 
It is straightforward to verify that with respect to the inner product (\ref{EQ:spin_inner_product}) we have  $J^\dagger_3=J_3$ and  $J_{\pm}^\dagger=J_{\mp}$.
 
Now from (\ref{EQ:coefficients}) and (\ref{EQ:z_basis})   we have  
\bea
\Psi(z)\equiv \brak{z}{\Psi}&=& \sum_m \brak{z}{j,m} \brak{j,m}{\Psi} \nonumber\\
&=& \sum_{m=-j}^j c_{j-m} z^{j+m}\nonumber\\
&=& P_\Psi(z).\nonumber
\eea
The Majorana polynomial is therefore precisely the coherent state wavefunction. For  each zero $z_i$  of the coherent state wavefunction, its {\it antipode\/} $-1/\bar z_i$  is the stereographic coordinate of the direction of one of the component  spin-$1/2$'s.  These antipodes are known as Majorana points \cite{aulbach,zimba}.

In the following  we sometimes find it  convenient  to use {\it normalized\/} coherent states
\be
\ket{z}_N = \frac{1}{(1+|z|^2)^j}\exp(\bz  J_+) \ket{j,-j},\quad _N\bra{z} = \frac{1}{(1+|z|^2)^j} \bra{j,-j}\exp(z J_-)=  \ket{z}_N^\dagger
\ee
 and the corresponding non-holomorphic coherent-state wavefunctions
 $$
  \psi(z, \bz)= _N\!\!\brak{z}{\psi}.
 $$
 In terms of these non-holomorphic wavefunctions the inner product becomes
 \be
\brak{\psi}{\chi}  = 
\frac {2j+1}{2\pi i } \int_{\mathbb C}  \frac{d\bz \wedge dz }{(1+| z|^2)^2}\,\overline{\psi(z,\bz)}\chi(z,\bz).
\label{EQ:newspin_inner_product}
\ee
The  integral  now involves only  the usual area form  on the  two-sphere expressed in stereographic coordinates. The normalized wavefunction 
$\psi(z,\bz)$ is therefore the probability amplitude for finding the spin-$j$ pointing in the direction $z$.

\begin{figure}
\includegraphics[scale=0.5]{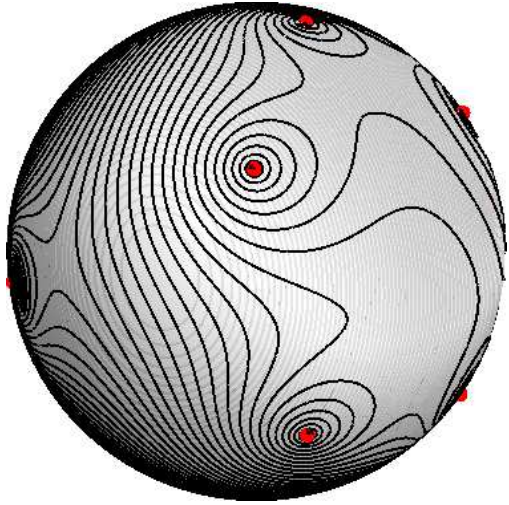}
\caption{A contour plot of  $\sqrt{|\psi(z, \bz)|}$ on the two-sphere for a typical  spin state $\ket{\psi}$. The location of its zeros is highlighted in red. The Majorana points of $\ket{\psi}$ are the antipodes  of these zeros.}
\label{FIG:generic-state}
\end{figure}

We can describe  a spin-$j$  state up to an overall phase by either specifying  the zeros $z_i$ of its spin coherent state wavefunction, or by specifying the Majorana points,  {\it i..e.\/}\  the directions on the Riemann sphere of  its component spin-$1/2$ factors. Because the  spin directions are the antipodes $w_i= -1/\bar z_i$  of the zeros,
 the   normalized spin-$1/2$ state  that points in the direction $w$ has  wavefunction
 \be
\psi_w(z,\bz)= \frac{1}{\sqrt{1+|z|^2}}\frac 1{\sqrt{1+|w|^2} }(1+z\bar w).
 \ee
This is just the wavefunction for the state $\ket{w}_N$.
Similarly we define
\bea
\ket{ \{ w_i \} }_N & \equiv & \ket{w_1,w_2,\ldots, w_{2j}}_N \nonumber \\
& = & \frac{1}{\mathcal{N}} \ket{w_1} \odot \ket{w_2} \odot \ldots \odot \ket{w_{2j}}
\eea
to be a normalized spin-$j$ state with its $2j$ spin-$1/2$ components pointing in the directions $w_i$. Here, ${\mathcal N}$ is a $w_i$ dependent 
normalization factor given by
\be
\mathcal{N}^2 = \brak{\{w_i\}}{\{w_i\}}= \frac{1}{(2j)!}\sum_{\pi\in\mathfrak{S}} 
	\prod_{i=1}^{2j} (1 + w_i \bar{w}_{\pi(i)}).
\ee
The corresponding wavefunction is 
 \be
_N\brak{z}{w_1,w_2,\ldots, w_{2j}}_N =(1+|z|^2)^{-j} \mathcal{N}^{-1} \prod_{i=1}^{2j} (1+ z \bar{w}_i).
 \label{EQ:majorana-wavefn}
 \ee

\section{Berry transport in the Majorana representation}
\label{SEC:berry}

Non-abelian Berry transport  is a generalization of the usual Berry adiabatic  transport  to the case of degenerate energy
eigenvalues \cite{wilczek-zee}.
Consider the adiabatic evolution of a state 
\be
\ket{\psi(t)}= a_n(t) \ket{n,t}
\ee
that 
lies  in  a subspace $V(t)$ spanned by a
orthonormal frame $\ket{n, t},n=1,\ldots, N$, of degenerate eigenstates of a  Hamiltonian $H(t)$. If $H(t)$ varies slowly, the resulting adaibatic evolution leads to the   coefficients being given by 
\be
a_n(t) = U_{nm}(t)a_m(0) \exp\left\{-i\int_0^t E(t) \dd t\right\},
\ee 
where $E(t)$ is the common eigenvalue, and 
\be
U_{nm}(t) =\left[\mathcal{P}\exp\left\{ - \int_0^t \mathcal{A}\right\}\right]_{nm}
\label{EQ:evolveU}
\ee
is a unitary matrix that generalizes the usual Berry phase.
The symbol $\mathcal{P}$ indicates a path-ordered integral
and  the non-abelian Berry
connection 
\be
\mathcal{A}_{nm} =
\eval{n,t}{\dd}{m, t} = - \bar{\mathcal{A}}_{nm}
\ee
is a skew-hermitian matrix-valued one-form. In the special case that degenerate  subspace is one-dimensional it reduces to the   abelian Berry connection.

We wish to find expressions for $\mathcal{A}_{nm}$ in the   Majorana parametrisation. The one dimensional (Abelian) case was obtained by Hannay \cite{hannay}.  He starts from    
\be
\dd\ket{\{w_i\}} = \frac{1}{(2j)!}\sum_{\pi \in {\mathfrak S}_{2j}}
\sum_i \ket{w_{\pi(1)}}\otimes\cdots(\dd\ket{w_{\pi(i)}})\cdots\otimes\ket{w_{\pi(2j)}}
\ee
where, from $\ket{w_i}= \bar w_i \ket{\!\uparrow} +\ket{\!\downarrow}$ we have $
\dd\ket{w_i} =  \dd\bar{w_i}\ket{\!\uparrow}$, 
and   
\be
 \mathcal{A} =i \,  \text{Im} \left(\frac{\eval{\{w_i\}}{\mathrm{d}}{\{w_i\}}}{\brak{\{w_i\}}{\{w_i\}}}\right).
\ee
He obtains 
\be
 {\mathcal A}= i\,  \text{Im}
  \frac{\sum_\pi\left[\prod_k(1+w_k \bar{w}_{\pi(k)}) 
  \sum_k (w_k \dd \bar{w}_{\pi(k)})/({1+w_k \bar{w}_{\pi(k))}})\right]}{\sum_\pi\left[\prod_k(1+w_k
  \bar{w}_{\pi(k)}\right]}. 
  \label{EQ:abelianphase}
\ee
For $j=1/2$, for example,  we have 
\bea
\mathcal{A} = i\, \text{Im}\left(\frac{\eval{w}{\mathrm{d}}{w}}{\brak{w}{w}}\right) =\frac 12 
\frac{w\text{d}\bar{w}-\bar{w} \dd w}{1+\bar{w}w}, \qquad 
\text{d}\mathcal{A} = \frac{\dd w\wedge\dd \bar{w}}{(1+\bar{w}w)^2}.
\eea
The expression for $\text{d}\mathcal{A}$ is $i$ times one-half the area form on the two-sphere, and so we find that familiar result that in a cyclic evolution the accumulated phase is one-half of the area swept out by the
spin. 

 For a general value of  $j$ consider the evolution of the  $E=m$ eigenstate of $H={\bf J}\cdot\hat  {\n}$ as the unit vector $\hat \n$ varies. In the Majorana
parametrisation, $j+m$ of the  $w_i$'s lie at the point  $w$ corresponding to the unit vector $\hat \n$ and $j-m$ of the $w_i$ lie at the antipodal point
$w'=-1/\bar{w}$. Thus equation (\ref{EQ:abelianphase}) becomes,
\bea
\mathcal{A}= \frac 12 (j+m)\frac{w\dd\bar{w}-\bar w \dd w}{1+\bar{w}w} +
\frac 12 (j-m)\frac{w'\dd\bar{w}'-\bar{w}'\dd w}{1+\bar{w}'w'}.
\eea
For a cyclic evolution the accumulated phase is $[(j+m)-(j-m)]/2=m$ times the area swept out by the point $w$ --- again a familiar result \cite{berry}.

Now  consider the non-abelian extension of Hanay's formula in which  Berry connection becomes a  one-form  $\mathcal{A}$ that  takes its values in the Lie algebra
$\mathfrak{u}(n)$. We can evaluate its  skew-hermitian matrix elements in the Majorana
parametrisation  by extending (\ref{EQ:abelianphase}). After some work we find that  the off diagonal elements of the connection are 
\bea
\mathcal{A}_{rs} &=& \frac{\eval{\{w^r_i\}}{\text{d}}{\{w^s_i\}}}
{\sqrt{\brak{\{w^r_i\}}{\{w^r_i\}}}\sqrt{\brak{\{w^s_i\}}{\{w^s_i\}}}}  \qquad
(r\neq s) \notag \\
&=&  \frac{\sum_\pi \prod_k(1+w^r_k \bar{w}^s_{\pi(k)}) \sum_k (w^r_k
\text{d} \bar{w}^s_{\pi(k)})/(1+w^r_k \bar{w}^s_{\pi(k)})}
{\sqrt{\sum_\pi\left[\prod_k(1+w^r_k
\bar{w}^r_{\pi(k)})\right]}\sqrt{\sum_\pi\left[\prod_k(1+w^s_k
\bar{w}^s_{\pi(k)})\right]}} .
\label{EQ:nonabelianphase1}
\eea
and the diagonal elements are
\bea
\mathcal{A}_{rr} &=&
\text{i}\frac{\text{Im}\eval{\{w^r_i\}}{\mathrm{d}}{\{w^r_i\}}}{\brak{\{w^r_i\}}{\{w^r_i\}}}
\notag \\
&=& \frac{\sum_\pi \text{Im}\left[\prod_k(1+w^r_k \bar{w}^r_{\pi(k)})
 \sum_k (w^r_k \dd \bar{w}^r_{\pi(k)})/(1+w^r_k
\bar{w}^r_{\pi(k)})\right]}{\sum_\pi\left[\prod_k(1+w^r_k
 \bar{w}^r_{\pi(k)})\right]}.
\label{EQ:nonabelianphase2}
\eea
The expressions in  equations (\ref{EQ:nonabelianphase1}) and  (\ref{EQ:nonabelianphase2}) are,  in general, rather complicated. Our interest, however, is in the evolution  of Kramers-degenerate pairs.  We therefore restrict ourselves to the time-reversal invariant hamiltonians, and to   half-integer $j$ for which the Kramers pair of states $\ket{\Psi}$ and ${\mathcal T}\ket{\Psi}$ are degenerate and mutually orthogonal. 

The time reversal operator on  a general spin state is  
\be
\mathcal{T} = \exp\{-i \pi J_2\}  {\mathcal K} 
\ee
where ${\mathcal K}$ denotes complex conjugation. On a spin-$1/2$ coherent state it acts as   
\be
\mathcal T \ket{w}= w\ket{-1/\bar w}
\ee
and this action extends to the general Majorana-parametrisation  tensor product state.

Using the states  $\ket{\Psi}$, ${\mathcal T}\ket{\Psi}$ as basis,  the Berry
connection becomes a two-by-two traceless skew-hermitian matrix,
\be
\mathcal{A} = \bpm \mathcal{A}_0& -\mathcal{A}_1 \\
\bar{\mathcal{A}_1}&-\mathcal{A}_0\epm.
\ee
The tracelessness of $\mathcal{A}$ is a
consequence of time-reversal; it implies that the diagonal Abelian phases for a Kramers pair
are complex-conjugates of each other. 

In the Majorana parametrisation  
\be
  \bar{\mathcal{A}_1} = \frac{\eval{\{w_i\}}{\text{d}}{\mathcal{T}\{w_i\}}}
{\sqrt{\brak{\{w_i\}}{\{w_i\}}}\sqrt{\brak{\mathcal{T}\{w_i\}}{\mathcal{T}\{w_i\}}}} =  
\frac{\sum_\pi\left[\prod_k(w_{\pi(k)}-w_k)\sum_k(\text{d}w_k)/(w_{\pi(k)}-w_k)\right]}
{\sum_\pi\left[\prod_k(1+\bar{w}_k w_{\pi(k)})\right]} \label{EQ:nab}
\ee
For eigenstates of $H={\bf J}\cdot \hat {\bf n}$ (\ref{EQ:nab}) simplifies considerably. Again, there are $j+m$ Majorana points $w$ and $j-m$ points at $w'=-1/\bar{w}$. The numerator contains the factor $\left[\prod_{k,k\neq j}(w_{\pi(k)}-w_k)\right]$, which can only be non-zero
for permutations that swap the $w$ and $w'$ except at most a single point that may map  to itself (recall that $2j$ is odd). As a consequence  the off-diagonal term is non-zero  only for $m=1/2$ states. When  $m\neq1/2$ the Berry transport  is purely Abelian, in the sense that  there exists a fixed basis whose  states are only multiplied by a phase for any  loop in Hamiltonian space.
	
Consider, for example  rotations of $\hat {\bf n}$  about the $z$ axis. We have $w_i\rightarrow
    w_i \exp{i\phi}$ and $\text{d}w_i=i w_i\text{d}\phi$ and equation \ref{EQ:nab})  becomes
\be
{\mathcal A}_1=\frac{\sum_P\left[\prod_k(w_{\pi(k)}-w_k)\sum_k{w_k}/(w_{\pi(k)}-w_k)\right]}
{\sum_P\left[\prod_k(1+\bar{w}_k w_{\pi(k)})\right]}\,\,i\text{d}\phi.
\ee
For a $m=1/2$ state with Majorana points at $w=\tan(\theta/2)\exp\{i\phi\}$ and
$w'=\cot(\theta/2)\exp\{-i\phi\}$ we get 
\bea
  \mathcal{A}_1 &=& \frac{(n+1)!(n+1)!}{n! (n+1)!} \frac{(1+w w')^{2n}/w'^{2n}}{(1+w w')^{2n+1}/(w
  w')^n} w (i \dd \phi) \nonumber \\
  &=& (n+1)\frac{w^n}{w'^n}\frac{1}{1+w w'} w  (i \dd \phi) \nonumber\\
  &=& \exp{[i(2n+1)\phi]}\frac{(n+1)}{2}\sin(\theta) (i \dd \phi)\label{EQ:mead1}
\eea
where $2n+1=2j$ is the total number of Majorana points. 
We may redefine the basis states so as to absorb the phase factor so that   
\be
  \mathcal{A}_1 = \frac{(n+1)}{2}\sin(\theta) (i \dd \phi) = \frac12 \left(j+\frac12\right)\sin(\theta) (i \dd \phi),
\ee
and this expression agrees with a result in    \cite{mead}.

\section{Kato's equation}
\label{SEC:kato}

The Berry-connection one-form allows us to adiabtically evolve states, but, because it involves continually  making a choice of basis,  it is not always the most convenient  method for numerical computation. An alternative  tool is provided by the  operator $P(t)$ that projects us into our  space $V(t)$ of     
 eigenstates of $H(t)$ with a common eigenvalue $E(t)$. This subspace evolves with  the  parameters in  $H(t)$, and the  Berry-transport condition \cite{wilczek-zee} is 
\be
P(t) \frac{\partial}{\partial t} \ket{\psi(t)} = 0.
\label{EQ:berrycond}
\ee
Intuitively, the Berry  condition says that we keep projecting the  state down into the evolving degenerate subspace. This process is completely equivalent to transport {\it via\/}  ${\mathcal A}_{nm}$.

Using the projection operators we can  approximate the adibatic evolution as
\be
\ket{\psi(t)}= P(t)P(t-\epsilon)P(t-2\epsilon)\cdots P(0) \ket{\psi(0)}.
\ee
For finite $\epsilon$, this   sequence of successive projections is not norm-preserving, but in the limit $\epsilon \to 0$ it  becomes so.

We now observe that 
 \bea
 \frac 1\epsilon\{\ket{\psi(t+\epsilon)}- \ket{\psi(t)}\} &=& \frac 1\epsilon\{P(t+\epsilon) -1\}P(t)P(t-\epsilon)\cdots P(0)\ket{\psi(0)}\nonumber\\
  &=& \frac 1\epsilon\{P(t+\epsilon)-P(t)\}P(t)P(t-\epsilon)\cdots P(0)\ket{\psi(0}.
 \eea
 Here we have used the projection-operator property $P(t)^2=P(t)$.
 On taking the  limit  $\epsilon \to 0$,  we are lead to    evolution equation
 \be
 \frac{\partial}{\partial t} \ket{\psi(t)} =\dot P(t) \ket{\psi(t)}.
 \label{EQ:evolution1}
 \ee
 Let us confirm  that this evolution satisfies equation (\ref{EQ:berrycond}):  from $P^2=P$ we have $\dot PP+P\dot P=\dot P$, and so
 \bea
 P \frac{\partial}{\partial t} \ket{\psi}_t &=&P\dot P(t) \ket{\psi(t)}\nonumber\\
 &=& (\dot PP-\dot P)\ket{\psi}\nonumber\\
 &=& (\dot P-\dot P)\ket{\psi}\nonumber\\
 &=&0.
 \eea
 In passing  from the second line to the third we have used $P\ket{\psi}=\ket{\psi}$   for $\ket{\psi}$ in the degenerate subspace $V(t)$.  
 
Although  equation   (\ref{EQ:evolution1}) is very simple, it has a formal drawback in that it does not  provide  a manifestly unitary state evolution. The  operator $P(t)$ is hermitian, and so therefore is $\dot P(t)$.  For unitary evolution we would expect the operator on the right hand side of (\ref{EQ:evolution1}) to be {\it skew\/} hermitian. We can, however,  equivalently write  (\ref{EQ:evolution1})  as 
 \be  
  \frac{\partial}{\partial t} \ket{\psi}_t =[\dot P(t), P(t)] \ket{\psi(t)}.
  \label{EQ:kato}
  \ee
 We have already seen that the the  added  term $P\dot P$ vanishes when acting on states in the degenerate eigenspace. Now the commutator of two hermitian operators  on the right-hand side {\it  is\/}   skew hermitian,  so this version of the  evolution equation  does provide  a manifestly  unitary evolution.  Equation (\ref{EQ:kato})  governing the  non-abelian Berry transport is known as {\it Kato's equation\/}  \cite{kato} after Tosio Kato who obtained it in 1950.

Now our interest is in the evolution of the spin in a molecule that is rotating in space. 
For a fixed orientation of the molecule, the spin  Hamiltonian is a   polynomial
\be
H({\bf J}) = \sum_{m=0}^{2j}  \sum_{i_1,\ldots i_m=1}^3 a_{i_1i_2\ldots i_m} J_{i_1}J_{i_2}\cdots J_{i_m}
\ee
in the space-fixed components $J_i$ of the spin angular momentum operator.  The effect of a real space rotation  $R$ on the molecule  is to  
is to change $J_i\to R^{-1}_{ij}J_j$ in this polynomial while keeping the coefficients $a_{i_1i_2\ldots i_m} $ fixed.   Now for  each $R$ there is a unitary operator $U[R]$  that implements the rotation in the $2j+1$ dimensional Hilbert space:
\be
R^{-1}_{ij}J_j = U[R]J_i U^{-1}[R],
\ee
so  as the molecule rotates we can write 
\be
H(t) = U[R(t)] H(0) U[R(t)]^{-1},
\ee
where  $U(t)=U[R(t)]$. (This $U(t)$ is not to be confused with the adiabatic evolution operator $U_{nm}(t)$ in equation (\ref{EQ:evolveU}).)   
The same operator evolves  the projection
\be
P(t) = U(t) P(0) U(t)^{-1},
\ee
and we   can  work in a frame rotating with the molecule  by setting $ \ket{\psi(t)} =  U(t) \ket{\phi(t)} $. In this  frame Kato's equation becomes
\be
\frac{\partial}{\partial t} \ket{\phi(t)} = -P(0) U^{-1} \dot{U} P(0) \ket{\phi(t)}. 
\label{EQ:kato-1}
\ee

The evolution given by (\ref{EQ:kato-1}) now all  takes place within $V(0)$ and  $\ket{\phi_t} = G(t) \ket{\phi_0}$, where the propagator $G(t)$ is  be the   time-ordered exponential
\bea
G(t) &=& \mathcal{P} \exp{\left(-\int_0^t P(0) K(t) P(0) \mathrm{d}s \right)}\nonumber\\
	& = & \lim_{\epsilon \to 0} \left\{P(0) (1 - \epsilon K(t)) P(0) (1 - \epsilon K(t-\epsilon))P(0) \cdots P(0) (1 - \epsilon K(0)) P(0)\right\},
\label{EQ:kato-time-slice}
\eea
where $K= U^{-1}\dot U$.
From (\ref{EQ:kato-time-slice}) we can see that $G(t) P(0) G(t)^{-1} = P(0)$.

\section{Action Principle and Equations of Motion}
\label{SEC:action}

The projection $P(0)$ in (\ref{EQ:kato-time-slice}) acts as  the identity operator on the subspace $V(0)$, and  is effectively the quantum evolution of a state restricted to $V(0)$.    We can therefore represent  (\ref{EQ:kato-time-slice}) by means of a Feynman path integral over trajectories in some suitable classical system.

As in the previous section we assume  time-reversal invariance and half-integer spin, so that by Kramers' theorem each energy level is doubly-degenerate and spanned by  $\ket{\phi(0)}$  and  $\mathcal{T}\ket{\phi(0)}$ (which we denote here as $\ket{\mathcal{T}\phi(0)}$). In such a space the  initial projection operator $P(0)$ is 
\be
P(0) = \ket{\phi(0)}\bra{\phi(0)} + \ket{\mathcal{T}\phi(0)}\bra{\mathcal{T}\phi(0)}.
\ee
Since $V(0)$ is two-dimensional, the natural path integral is the spin-coherent-state path integral for ${\rm SU}(2)$. 

To obtain the path integral we let  $\lambda \in \mathbb{C}P^1$ be the stereographic complex coordinate of a point on the Riemann sphere and set 
\bea
\ket{\lambda}_N & = & (1+\lambda \bar \lambda)^{-1/2} (\ket{\mathcal{T} \phi(0)} + \bar \lambda \ket{\phi(0)}) \\
\bra{\lambda}_N & = & (1+\lambda \bar \lambda)^{-1/2} (\bra{\mathcal{T} \phi(0)} + \lambda \bra{\phi(0)}).
\eea
These are analogous to the spin-$1/2$ coherent states in which  $\ket{\mathcal{T}\phi(0)}$ and $\ket{\phi(0)}$ are  playing the roles of the spin down and spin up states respectively.

We  now write the projection operator $P(0)$ as the  (overcomplete) resolution
\be
P(0) = \frac{2}{4\pi}\int_{\mathbb{C}P^1} \frac{2 \, \mathrm{d}\bar\lambda \wedge \mathrm{d}\lambda}{i(1+\lambda\bar\lambda)^{2}} {}_N\ket{\lambda}\bra{\lambda}_N,
\label{EQ:overcomplete-lambda}
\ee
and with it construct, by time-slicing with $P(0)$, what is essentially a spin-$1/2$ ``coherent" state path-integral \cite{stone-park-garg} containing the action 
\be
S[\lambda, \bar \lambda]] = \int_0^T \, \mathrm{d}t 
\left( \frac{1}{2} \frac{\bar \lambda \dot{\lambda} - \dot{\bar \lambda} \lambda}{1 + \lambda \bar \lambda} - \mathcal{A}_0(\lambda, \bar \lambda) 
\right),
\ee
with $\mathcal{A}_0 \equiv \eval{\lambda}{K}{\lambda}$.

We might  attempt to evaluate this path integral by some technique.  However, when the Hamiltonian term is {\it linear\/} in the Lie algebra generators, as it is in this  case, then  we know that the {\it classical\/}  equations of motion  give the exact quantum evolution (See for example \cite{onofri88,stone-note}).
 
These equations are
\bea
\frac{d \lambda}{d t} & = & (1 + \lambda \bar \lambda)^2 \frac{\partial \mathcal{A}_0}{\partial \bar \lambda} \\
\frac{d \bar \lambda}{d t} & = & -(1 + \lambda \bar \lambda)^2 \frac{\partial \mathcal{A}_0}{\partial \lambda} \label{EQ:semiclassical-lambda}
\eea
Now, from the construction of $\ket{\lambda}$, we can write
\be
(1+\lambda \bar \lambda) \frac{\partial}{\partial \bar \lambda} \eval{\lambda}{X}{\lambda} = 
-\eval{\lambda}{X \, \mathcal{T}}{\lambda} \label{EQ:lambda-property-1}
\ee
for any operator $X$. In particular we can  rewrite (\ref{EQ:semiclassical-lambda}) as
\bea
\frac{d \lambda}{d t} & = & -(1+\lambda \bar \lambda) \mathcal{A}_1 \\
\frac{d \bar \lambda}{d t} & = & -(1+\lambda \bar \lambda) \bar{\mathcal{A}}_1 \label{EQ:semiclassical-lambda-2}
\eea
where $\mathcal{A}_1(\lambda,\bar \lambda)\equiv\eval{\lambda}{K\,\mathcal{T}}{\lambda}$. We now know the time evolution of $\lambda(t)$, and hence the  time evolution of $\ket{\phi(t)}$.

To  see how changes in $\lambda$ and $\bar \lambda$ translate into changes in the location of the zeros $z_i$, or equivalently the Majorana points $w_i= -1/\bar z_i$ we observe that 
\be
\ket{\lambda}_N = e^{i \gamma} \ket{\{w_i\}}_N\label{EQ:EOM-ket-relation}
\ee
for some phase $\gamma$, and 
so
\bea
\psi(z,\bar\lambda) &=& e^{i \gamma} \mathcal{N}^{-1} (1+|z|^2)^{-j}
	\prod_{i=1}^{2j} (1 + z \bar w_i)\nonumber\\
	&=&  Q\prod_{i=1}^{2j} (1 + z \bar w_i),
	 \label{EQ:EOM-wavefn}
\eea
where $Q = e^{i \gamma} \mathcal{N}^{-1} (1+|z|^2)^{-j}$.  The last form reminds us that 
the roots of the wavefunction $\psi(z,\bar\lambda)$ are at $z_i = -1 / \bar w_i$.
Assume for now that the zeros are all non-degenerate --- that is, $z_i \neq z_j$ if $i \neq j$.  Consider a root $z_k(\bar\lambda)$ of $\brak{z}{\lambda}$,  and recall that the zeros of  $\brak{z}{\lambda}$  depend  only on  $\bar\lambda$ while those of  $\brak{\lambda}{z}$ depend only on $\lambda$.  Let  $\bar\lambda\to\bar\lambda+\delta\bar\lambda$ and expand $z_k(\bar\lambda+\delta\bar\lambda)$ to first order as
\be
z_k(\bar\lambda+\delta\bar\lambda) = z_k(\bar\lambda) + \frac{d z_k}{d \bar\lambda} \delta\bar\lambda,
\ee
For $z_k(\lambda+\delta\bar\lambda)$ to remain a root, we must  have
\be
\psi(z_k + \frac{d z_k}{d \bar\lambda} \delta\bar\lambda, \, \bar\lambda+\delta\bar\lambda) = 0.
\label{EQ:EOM-ND-1}
\ee
Expanding (\ref{EQ:EOM-ND-1}) to first order and using  $\psi(z_k,\bar\lambda)=0$, we find  \cite{ellinas,leboeuf}
\be
\frac{d z_k}{d \bar\lambda} = -\frac{\partial\psi}{\partial \bar\lambda} \left(\frac{\partial \psi}{\partial z}\right)^{-1} \bigg|_{z_k}.
 \label{EQ:EOM-ND-2}
\ee
The denominator of (\ref{EQ:EOM-ND-2}) is obtained by straightforward differentiation of (\ref{EQ:EOM-wavefn}) to be
\be
\frac{\partial \psi}{\partial z}\bigg|_{z_k} = -Q \, \bar{w}_k^{2-2j} \prod_{i \neq k}(\bar{w}_k - \bar{w}_i)\label{EQ:EOM-ND-denom},
\ee
and the numerator is
\be
\frac{d\psi}{d \bar\lambda}\bigg|_{z_k} = -(1 + |\lambda|^2)^{-1} e^{-2i \gamma} Q\, \bar{w}_k^{-2j} \prod_i(1 + w_i \bar{w}_k).\label{EQ:EOM-ND-numer}
\ee
Combining equations (\ref{EQ:EOM-ND-1}), (\ref{EQ:EOM-ND-denom}), (\ref{EQ:EOM-ND-numer}), and using that
\be
\frac{d z_k}{d \bar\lambda} = \frac{d}{d \bar\lambda} \left(-\frac{1}{\bar {w}_k} \right) = 
	-\frac{1}{\bar{w}^2_k} \frac{d \bar{w}_k}{d \bar\lambda},
\ee
we arrive at differential equations for the Majorana points $w_k$ in terms of $\lambda$,
\bea
(1+\lambda\bar\lambda)\frac{d w_k}{d \lambda} & = & e^{2 i \gamma} (1 + w_k \bar{w}_k) \prod_{i \neq k} \frac{1 + \bar{w}_i w_k}{w_k - w_i} ,\\
(1+\lambda\bar\lambda)\frac{d \bar{w}_k}{d \bar\lambda} & = & e^{-2 i \gamma} (1 + w_k \bar{w}_k) \prod_{i \neq k} \frac{1 + w_i \bar{w}_k}{\bar{w}_k - \bar{w}_i}. \label{EQ:EOM-ND-lambda}
\eea
Now 
\be
\frac{d w}{d t} = \frac{d w}{d \lambda} \frac{d \lambda}{d t} = \frac{d w}{d \lambda} (1+|\lambda|^2) \mathcal{A}_1,
\ee
where we have used (\ref{EQ:semiclassical-lambda-2}),
and 
\be
\mathcal{A}_1 = {}_N\eval{\lambda}{U^{-1}\dot U\,\mathcal{T}}{\lambda}_N = e^{-2 i \gamma} {}_N\eval{\{w\}}{U^{-1}\dot U \,\mathcal{T}}{\{w\}}_N.
\ee
The phase $\gamma$   cancels leaving us with 
\bea
\frac{d w_k}{d t} & = & _N\eval{\{w\}}{U^{-1}\dot U \, \mathcal{T}}{\{w\}}_N (1+w_k\bar{w}_k)
	\prod_{i \neq k} \frac{1 + \bar{w}_i w_k}{w_k - w_i}, \nonumber\\
\frac{d \bar{w}_k}{d t} & = & {}_N\overline{\eval{\{w\}}{U^{-1}\dot U  \, \mathcal{T}}{\{w\}}}_N (1+w_k\bar{w}_k)
	\prod_{i \neq k} \frac{ 1 + w_i \bar{w}_k} {\bar{w}_k - \bar{w}_i} .\label{EQ:EOM-ND-t}
\eea

The equations  (\ref{EQ:EOM-ND-t}) are quite intricate. To gain some insight,  consider some of their properties.
We begin by locating possible fixed points of (\ref{EQ:EOM-ND-t}). If $\dot w_k$ is to be zero there are two possibilities: either the the product $\prod_{i \neq k}(1+\bar{w}_i w_k)(w_k-w_i)^{-1}$ must vanish, or the factor  ${}_N\eval{\{w\}}{U^{-1}\dot U \, \mathcal{T}}{\{w\}}_N$  must vanish.
We focus on the first possibility.
For this to occur, one of the Majorana points $w_i$ must become antipodal to $w_k$ ---  that is  there is a $w_{k'}(t)$ such that
\be
w_{k'}(t) = -\frac{1}{\bar{w}_k(t)}.
\ee
At that instant, $\dot{w}_k = 0$. But observe that this also hold true under $k\to k'$, and $\dot{w}_k = 0$ as well. Consequently, both $w_k$ and $w_{k'}$ remain static for all $t$. As such, antipodal Majorana points are always fixed.

We can understand this by noting that antipodal Majorana points are actually Majorana points that are common to both $\ket{\phi_t}$ and $\ket{\mathcal{T}\phi_t}$. Suppose the Majorana points of $\ket{\phi_t}$ are given by
\be
a_1, -1/\bar{a}_1, \ldots, a_r, -1/\bar{a}_r, \xi_1, \ldots, \xi_p \quad\quad (2r+p = 2j)\label{EQ:free-mp-convention}
\ee
where the $a_i$'s come in antipodal pairs. Since $\mathcal{T}$ takes the Majorana points to their antipodes, the $a_i$'s are preserved; in other words, the Majorana points of $\ket{\mathcal{T}\phi}$ are
\be
a_1, -1/\bar{a}_1, \ldots, a_r, -1/\bar{a}_r, -1/\bar{\xi}_1,\ldots,-1/\bar{\xi}_p.
\ee
Note that  $\ket{\phi_t}$ is a linear combination of $\ket{\phi_0}$ and $\ket{\mathcal{T}\phi_0}$,  when $\ket{\phi_t}$ and $\ket{\mathcal{T}\phi_0}$ have any Majorana points in common, then  $\ket{\phi_t}$ will also share these Majorana points.

To summarize:   if a pair of Majorana points start out being antipodal, then they  remain fixed  at all times. In future  we will factor out these fixed points, and refer to the  non-fixed Majorana points as ``free" or ``dynamical".

Since the fixed Majorana points do not participate in the dynamics, the requirement in equation (\ref{EQ:EOM-ND-t}) of non-degenerate Majorana points can be relaxed to only requiring that none of the free Majorana points ever coincide. This includes states whose fixed antipodal pairs occure more than once (for example, the $\ket{j,1/2}, \ket{k,-1/2}$ states).

Writing $\xi_i$ for the free Majorana points and $a_i, -1/\bar{a}_i$ for the fixed antipodal pairs as in (\ref{EQ:free-mp-convention}), we find that 
\bea
\frac{d \xi_k}{d t} & = & 
	_N\eval{\{a\},\{\xi\}}{U^{-1}\dot U \, \mathcal{T}}{\{a\},\{\xi\}}_N 
	(1+\xi_k\bar{\xi}_k) \, \eta
	\prod_{\substack{1\leq i \leq p\\ i\neq k}} 
	\frac{1 + \bar{\xi}_i \xi_k}{\xi_k - \xi_i} \nonumber \\
\frac{d \bar{\xi}_k}{d t} & = & 
	_N\overline{\eval{\{a\},\{\xi\}}{U^{-1}\dot U \, \mathcal{T}}{\{a\},\{\xi\}}}_N 
	(1+\xi_k\bar{\xi}_k) \, \bar\eta
	\prod_{\substack{1\leq i \leq p\\ i\neq k}} 
	\frac{ 1 + \xi_i \bar{\xi}_k} {\bar{\xi}_k - \bar{\xi}_i} \label{EQ:EOM-ND-t-free}
\eea
Here  $\eta = \prod_{i=1}^{r} (-\bar{a}_i/a_i)$ is  an $a$-dependent overall phase, and  $\ket{\{a\},\{\xi\}}$ denotes  a state in the Majorana representation with fixed Majorana points $a, -1/\bar{a}$, and free Majorana points $\xi$. Observe that (\ref{EQ:EOM-ND-t-free}) is well behaved even if $\xi_k$ coincides with one of the $a$'s,  but becomes  singular if any $\xi_k$ approaches another free $\xi_{k'}$.

An especially interesting  case occurs when only one Majorana point is free and the rest  compose  antipodal pairs. In this case the equations of motion reduce to
\bea
\frac{d \xi}{d t} & = & {}_N\eval{\{a\},\xi}{U^{-1}\dot U \, \mathcal{T}}{\{a\},\xi}_N 
				(1+\xi \bar{\xi}) \, \eta \\
\frac{d \bar\xi}{d t} & = &{}_N \overline{\eval{\{a\},\xi}{U^{-1}\dot U\, \mathcal{T}}{\{a\},\xi}}_N 
				(1+\xi_k\bar{\xi}_k) \, \bar\eta \label{EQ:EOM-one-free}
\eea

What is interesting here is that (\ref{EQ:EOM-one-free}) can be derived from  an action 
\be
S[\xi,\bar \xi]  = \int_0^T \mathrm{d}s \left(
	\frac{1}{2}\frac{\bar\xi \dot{\xi} - \xi \dot{\bar\xi}}{1+\xi \bar\xi} - 
	{}_N\eval{\{a\},\xi}{U^{-1}\dot U}{\{a\},\xi}_N \right).
\ee
The semiclassical equations of motion obtained from this action are 
\bea
\frac{d \xi}{dt} & = & -(1+\xi\bar\xi)^2 \frac{\partial}{\partial \bar\xi}{}_N \eval{\{a\},\xi}{U^{-1}\dot U}{\{a\},\xi} _N\nonumber\\
\frac{d \bar{\xi}}{dt} & = & {\phantom-}(1+\xi\bar\xi)^2 \frac{\partial}{\partial \bar\xi}{}_N \eval{\{a\},\xi}{U^{-1}\dot U}{\{a\},\xi}_N ,\label{EQ:EOM-one-free-2}
\eea
At first glance equations (\ref{EQ:EOM-one-free}) and (\ref{EQ:EOM-one-free-2})  look different.  That they are, equivalent however, can be verified by using equation (\ref{EQ:lambda-property-1})

\section{Illustration}
\label{SEC:illustration}

As an application of our  formalism,  consider  the family of molecular  magnets Mn4. The effective Hamiltonian for the $j=9/2$ molecular spin in the absence of an external field can be written as 
\be
\mathcal{H} = k_x J_x^2 + k_y J_y^2 + k_z J_z^2,
\ee
together with small quartic terms that we will ignore.   (The small field-coupling between the spins on different molecules will also produce a term linear in $J$ --- and therefore time-reversal symmetry breaking --- which may be significant. We are, however, also going to ignore this  for the moment.)   
 Following \cite{mettes}, we adopt the values  $k_x = -k_y = 0.0243 \mathrm{K}$, $k_z = -0.59 \mathrm{K}$.

\begin{figure}
\begin{center}
{\includegraphics[scale=0.35]{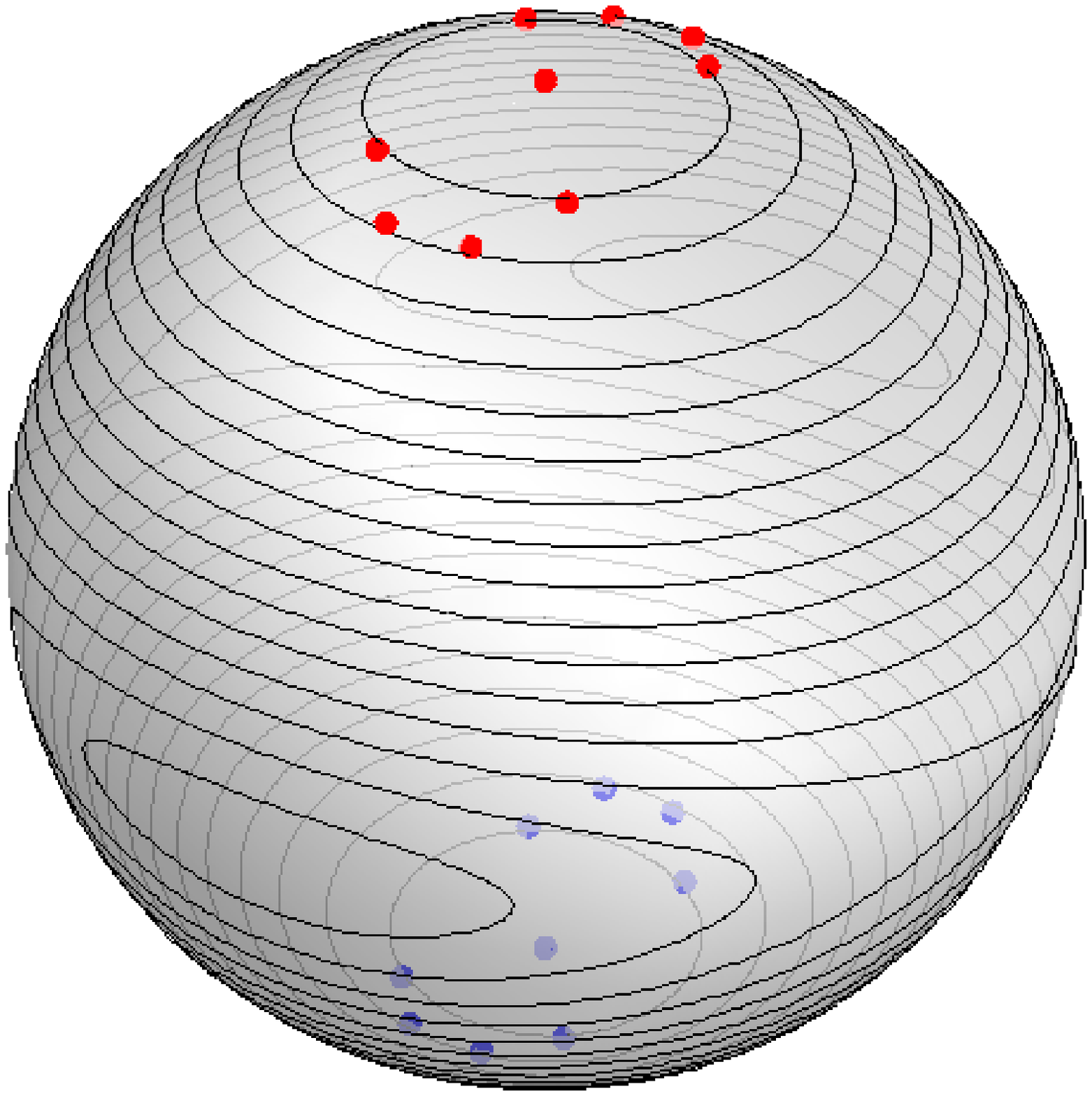}}
\hspace {15pt}
{\includegraphics[scale=0.35]{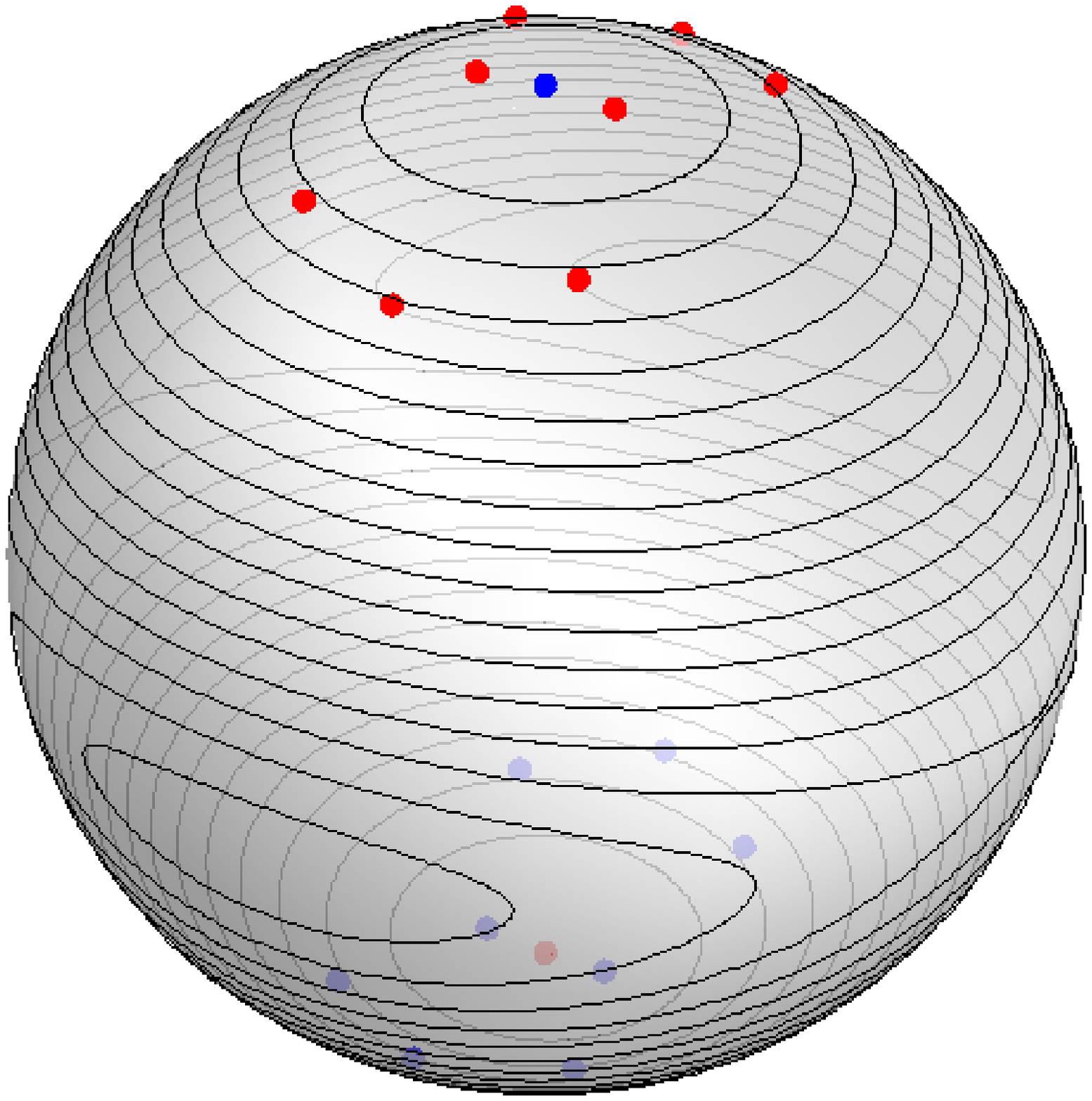}}\\
\vspace{12pt}
{\includegraphics[scale=0.35]{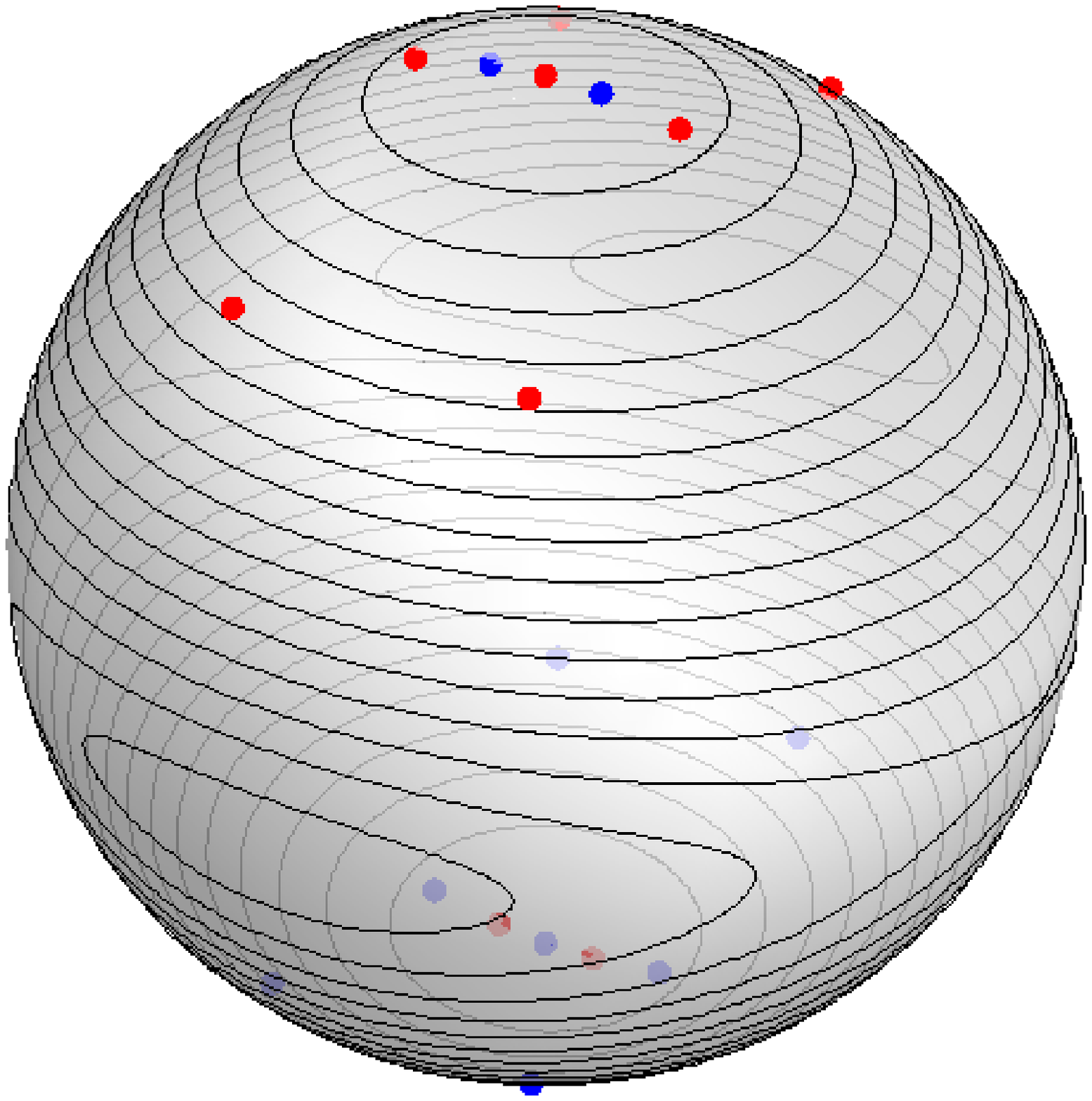}}
\hspace{15pt}
{\includegraphics[scale=0.35]{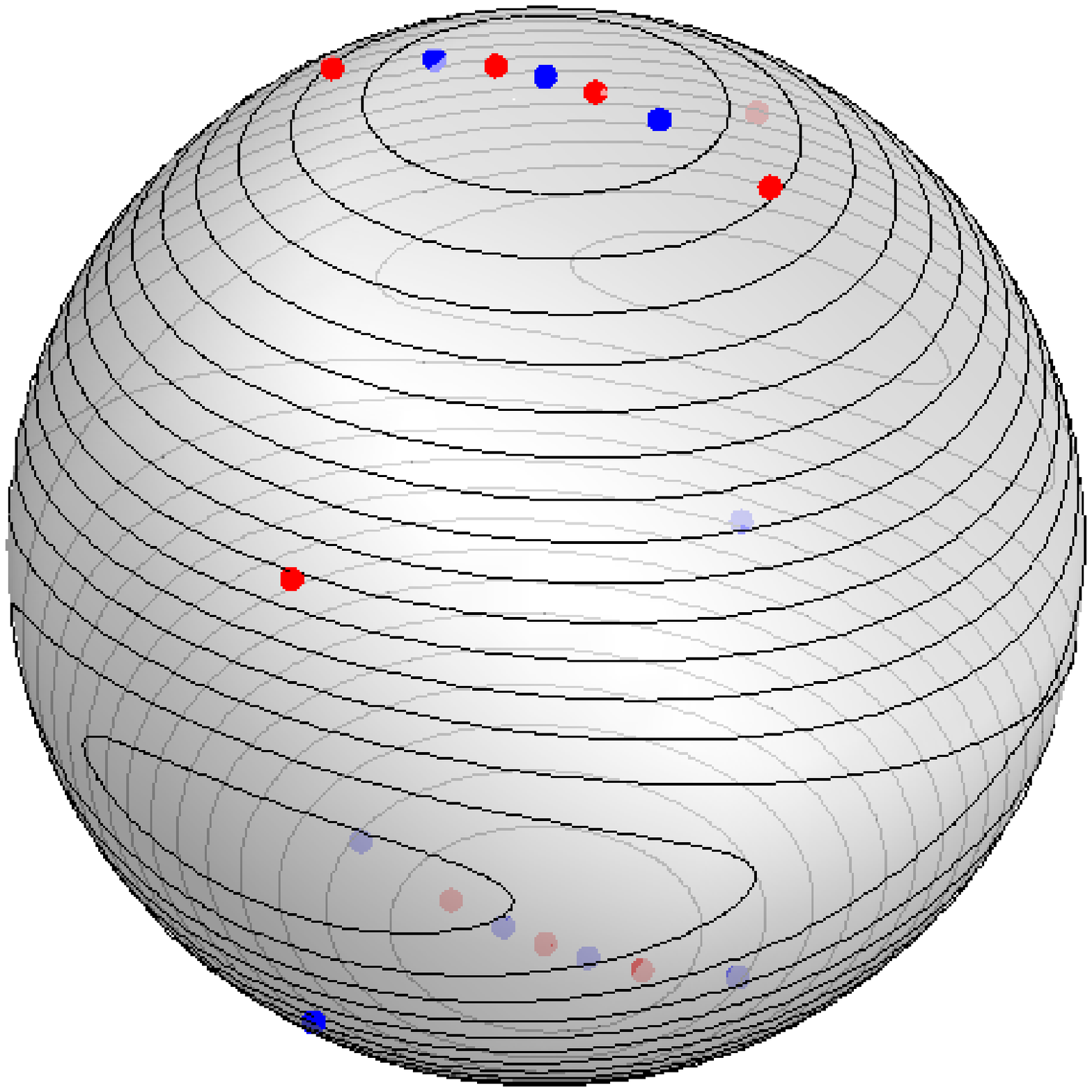}}\\
\vspace{12pt}
{\includegraphics[scale=0.35]{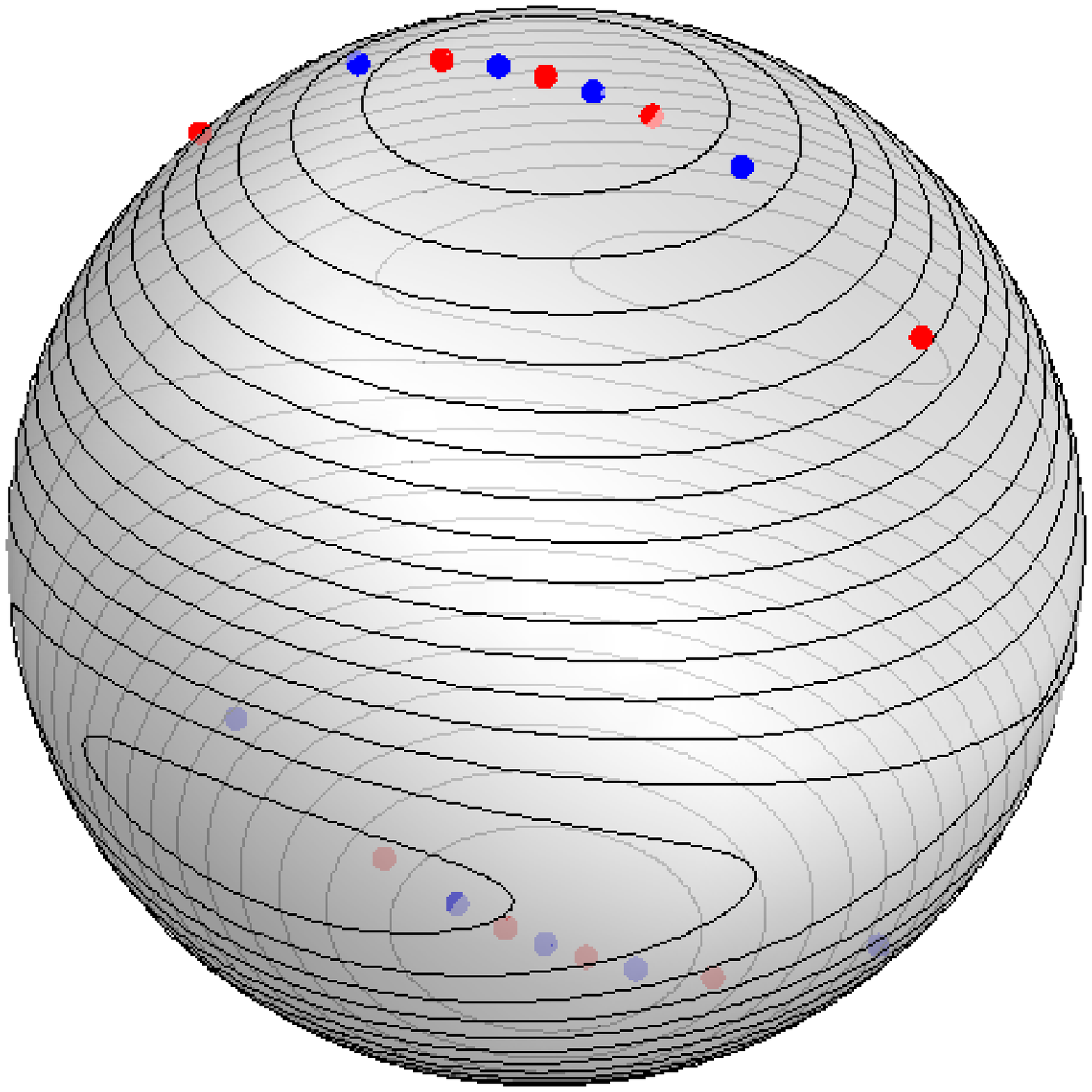}}
\end{center}
\caption{The zeros of the eigenstate wavefunctions for $\mathcal{H}$ in absence of an applied field, for which the expectation $\vev{\bold{J}}$ is largest about the z-axis. Blue and red denote the zeros of the eigenstate and its Kramers pair, repsectively. In order from left to right and top to bottom, we have (1) ground state, (2) first excited state, (3) second excited state, (4) third excited state, (5) highest state. In the background we have plotted the energy contours of corresponding semiclassical Hamiltonian.}
\label{FIG:eigenstates-zkett}
\end{figure}

First, we look at the configurations of the zeros for the eigenstate wavefunctions of $\mathcal{H}$, for which there are a total of $5$ doubly-degenerate levels when $j=9/2$. Due to the relatively large quadrupole moment $k_z$ of the $z$ axis, the Hamiltonian is close to that of $J_z^2$, and as a result the zeros of the wavefunctions will want to cluster about the poles (see figure \ref{FIG:eigenstates-zkett}). But unlike zeros of $\brak{z}{j,m}$ which simply condense at the poles, the small $J_x^2$ and $J_y^2$ terms will cause some repulsion between them. In the background we also display the contours of the ``classical'' potential landscape defined by the expectation $h(z,\bar z)\stackrel{\rm def}{=}  \vphantom{}_N\eval{z}{H}{z}_N$.

\begin{figure}
\begin{center}
{\includegraphics[scale=0.35]{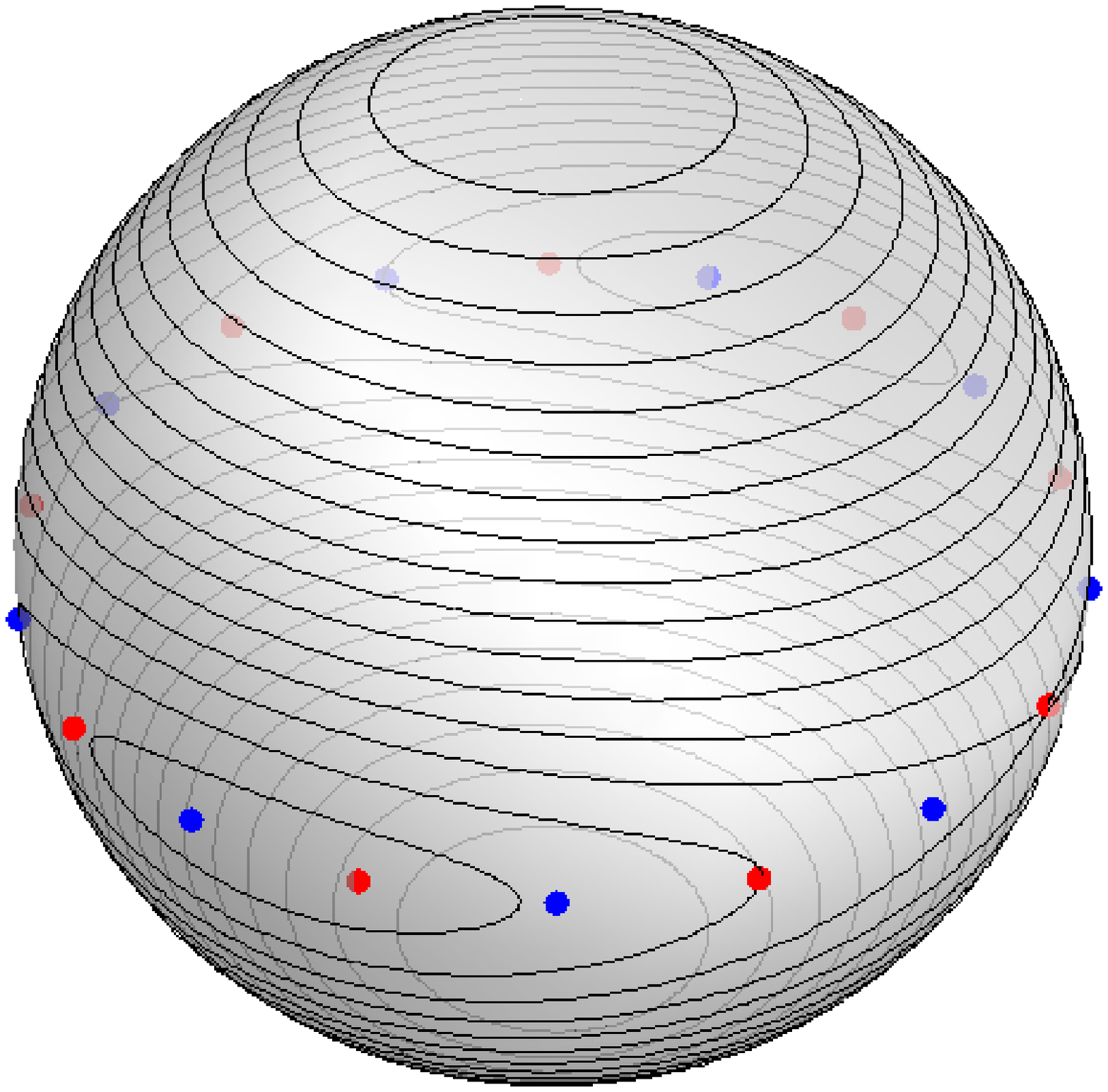}}
\hspace {15pt}
{\includegraphics[scale=0.35]{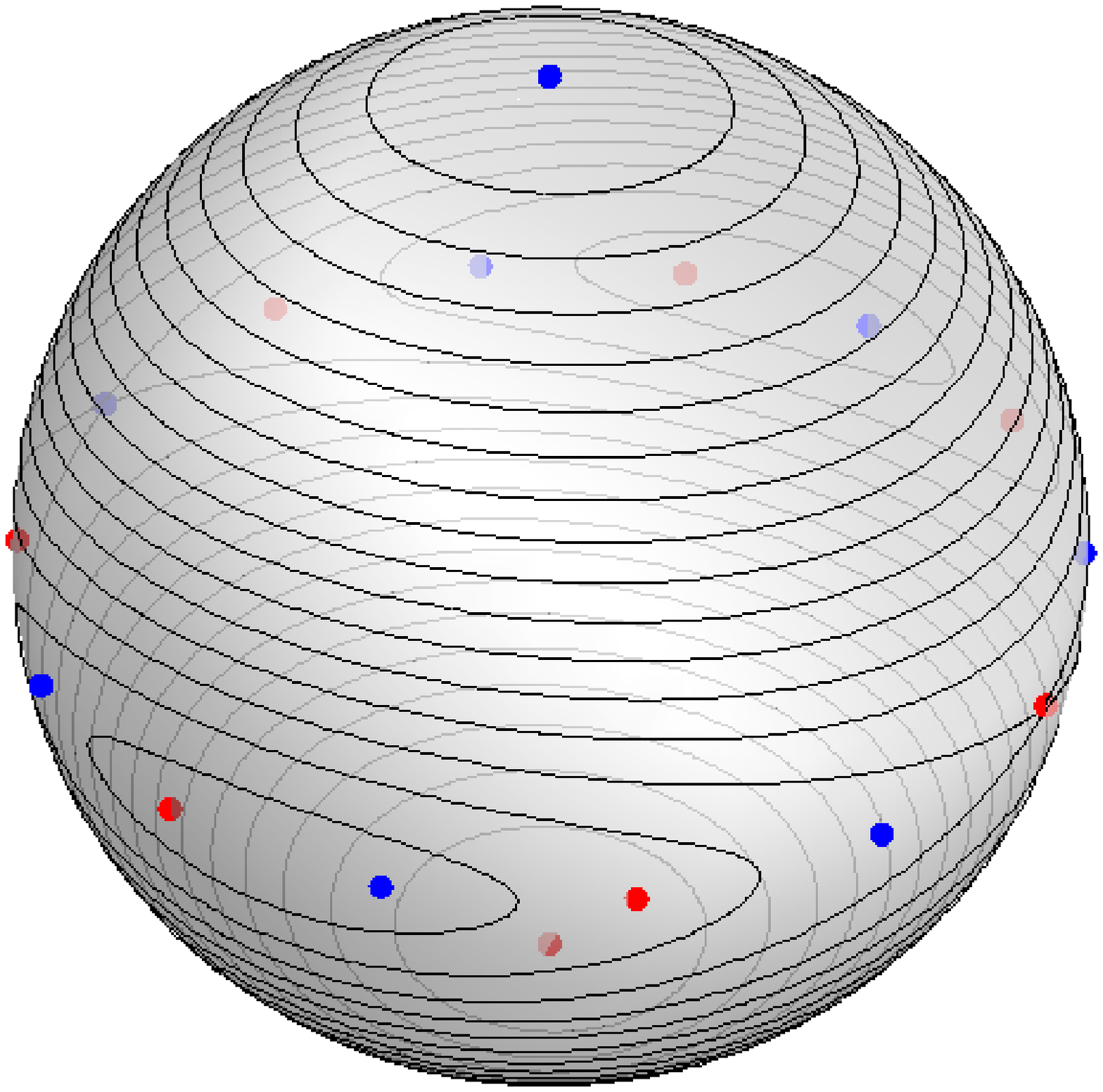}}\\
\vspace{12pt}
{\includegraphics[scale=0.35]{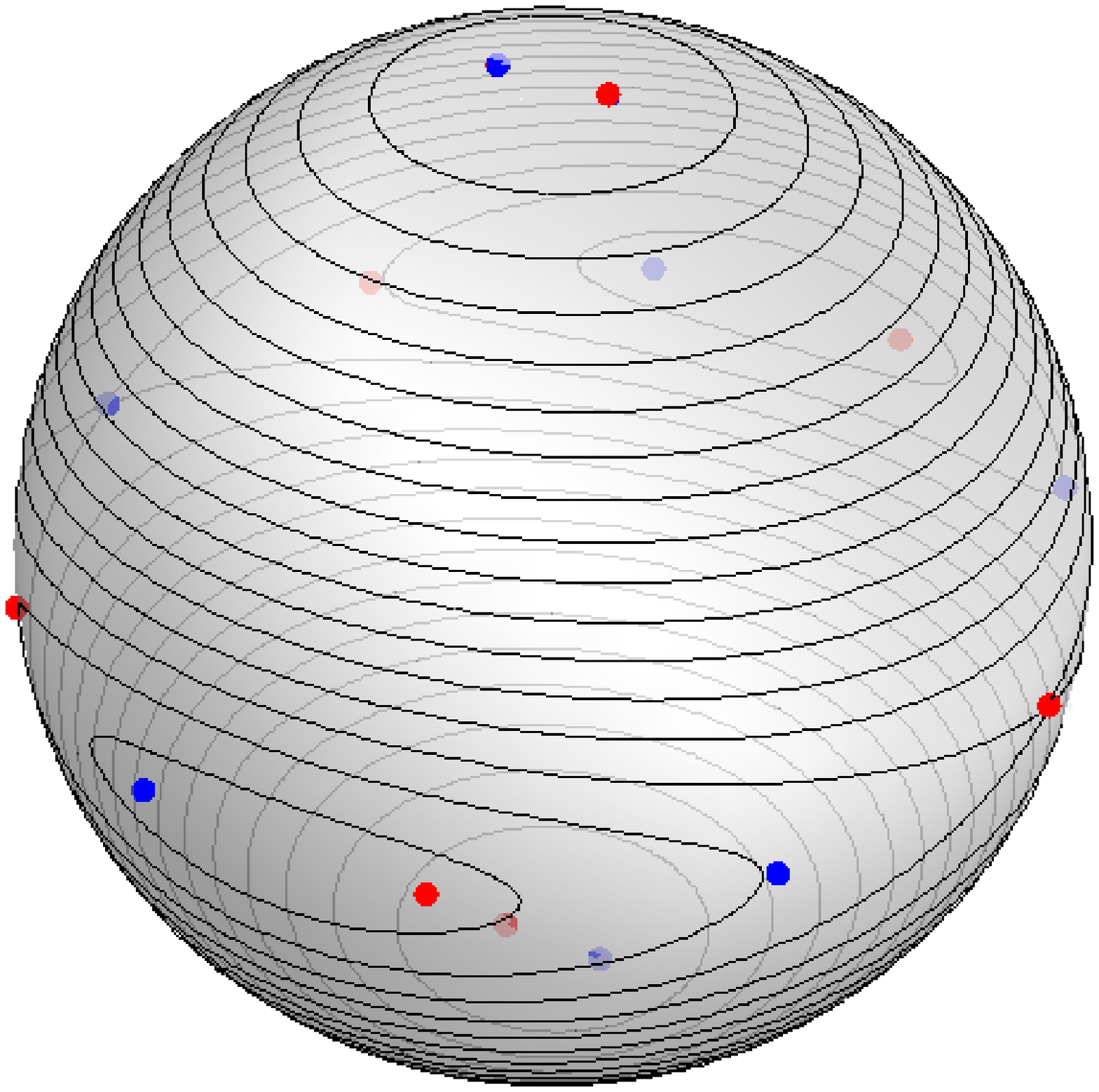}}
\hspace{15pt}
{\includegraphics[scale=0.35]{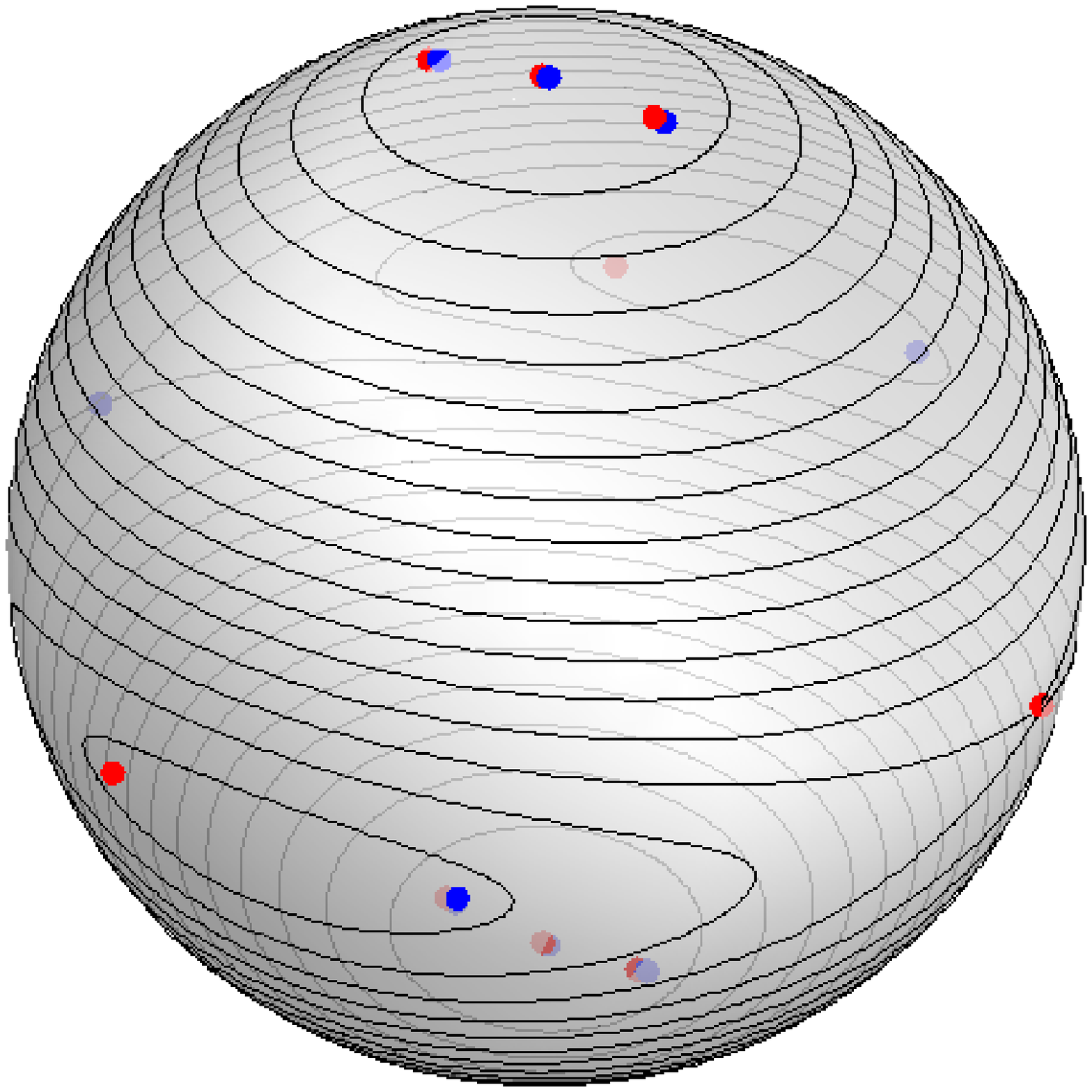}}\\
\vspace{12pt}
{\includegraphics[scale=0.35]{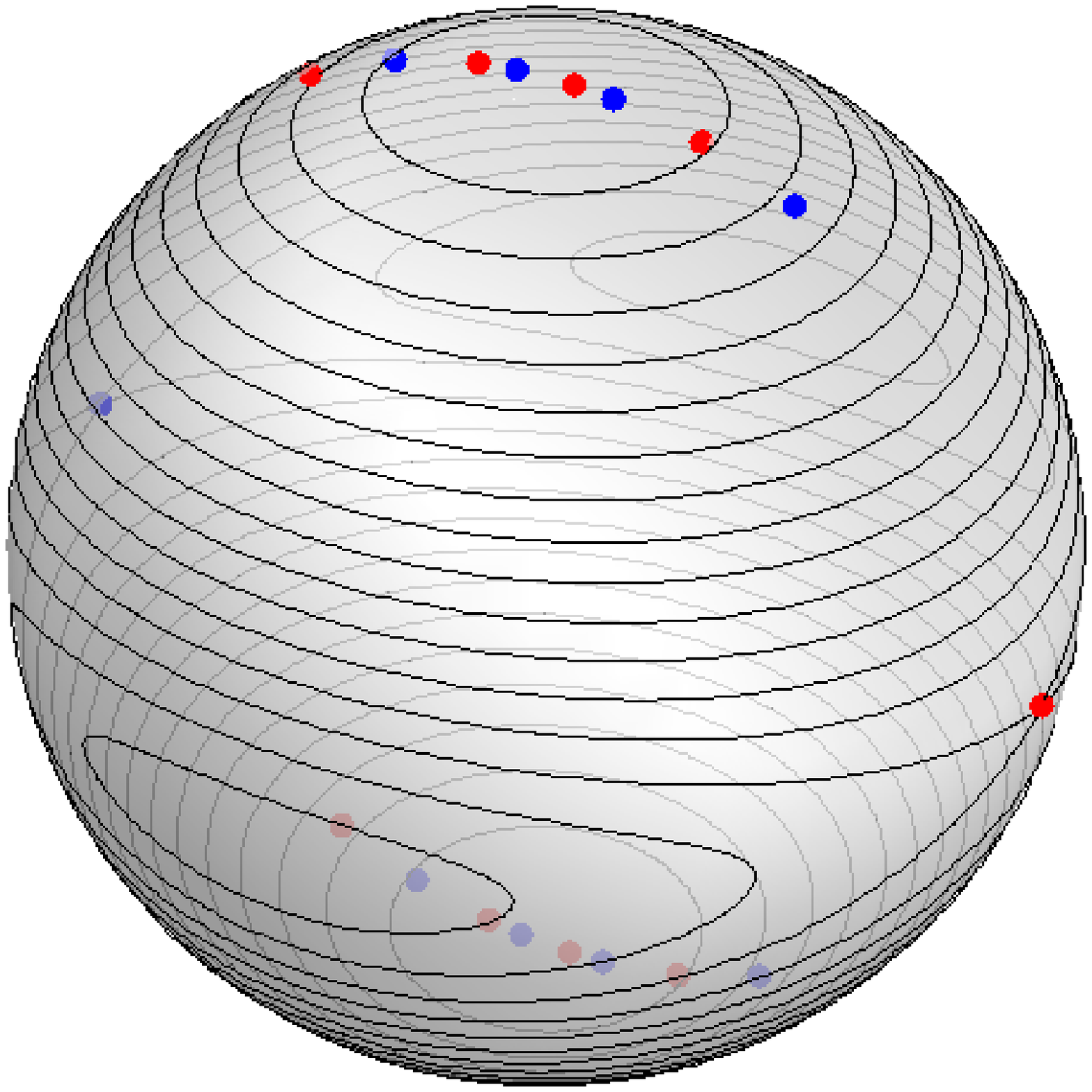}}
\end{center}
\caption{The zeros of the eigenstate wavefunctions for $\mathcal{H}$ in absence of an applied field, for which the initial states are prepared by polarizing along the $\hat{\bold{x}}$. Note that in the first, second, and third excited states the zeros about the poles actually very closely overlap --- for example the blue dot in the north pole for the first excited state is actually a pair of blue and red dots.}
\label{FIG:eigenstates-xkett}
\end{figure}

Due to the double degeneracy however, there is a freedom in the choice of the basis of the eigenspace. One particularly illuminating selection is made by putting in a small transverse magnetic field in the $\hat{\bold{x}}$ direction and relaxing the field strength to zero. In   figure \ref{FIG:eigenstates-xkett} we see that for the ground states the zeros tend to distribute themselves along flow lines $\nabla h(z,\bar z)$ in a manner such that the two ground states have all their zeros distributed about regions of maximum energy, while for the first excited states a pair of zeros for each state have relocated themselves close to the minima, for the second excited state an additional pair, etc. In addition, for the first, second, and third excited states we find that the zeros about the poles of one state very closely overlap with that of its Kramers pair, and this can have an effect when considering the dynamics their dynamics under adiabatic evolution. We remark however that for this choice of states the magnetization $\vev{\bold{J}}$ becomes significantly reduced, and may consequently be very difficult to measure.

Now we consider adiabatic holonomy arising from rotations about a fixed axis $\hat{\bold{n}} = \sin \eta \cos \gamma\, \hat{\bold{x}} + \sin \eta \sin \gamma\,  \hat{\bold{y}} + \cos \eta\,  \hat{\bold{z}}$ at an angular velocity of $2 \pi / T$. For such rotations we are guaranteed to have the state return to itself (in the corotating frame) after some time $\tau$, which need not coincide in general with the period of rotation $T$; the former is gotten by decomposing $P U^{-1} \dot U P$ into sigma matrices $x_i \sigma_i$ in the eigenstate basis, for which the period is then found as $2 \pi / |x|$. Fixed-axis rotations have been considered before by \cite{mead}, but their attention is restricted to eigenstates of isotropic Hamiltonians of the form $\hat{\bold{n}}\cdot \bold{J}$. 

\begin{figure}
\begin{center}
{\includegraphics[scale=0.37]{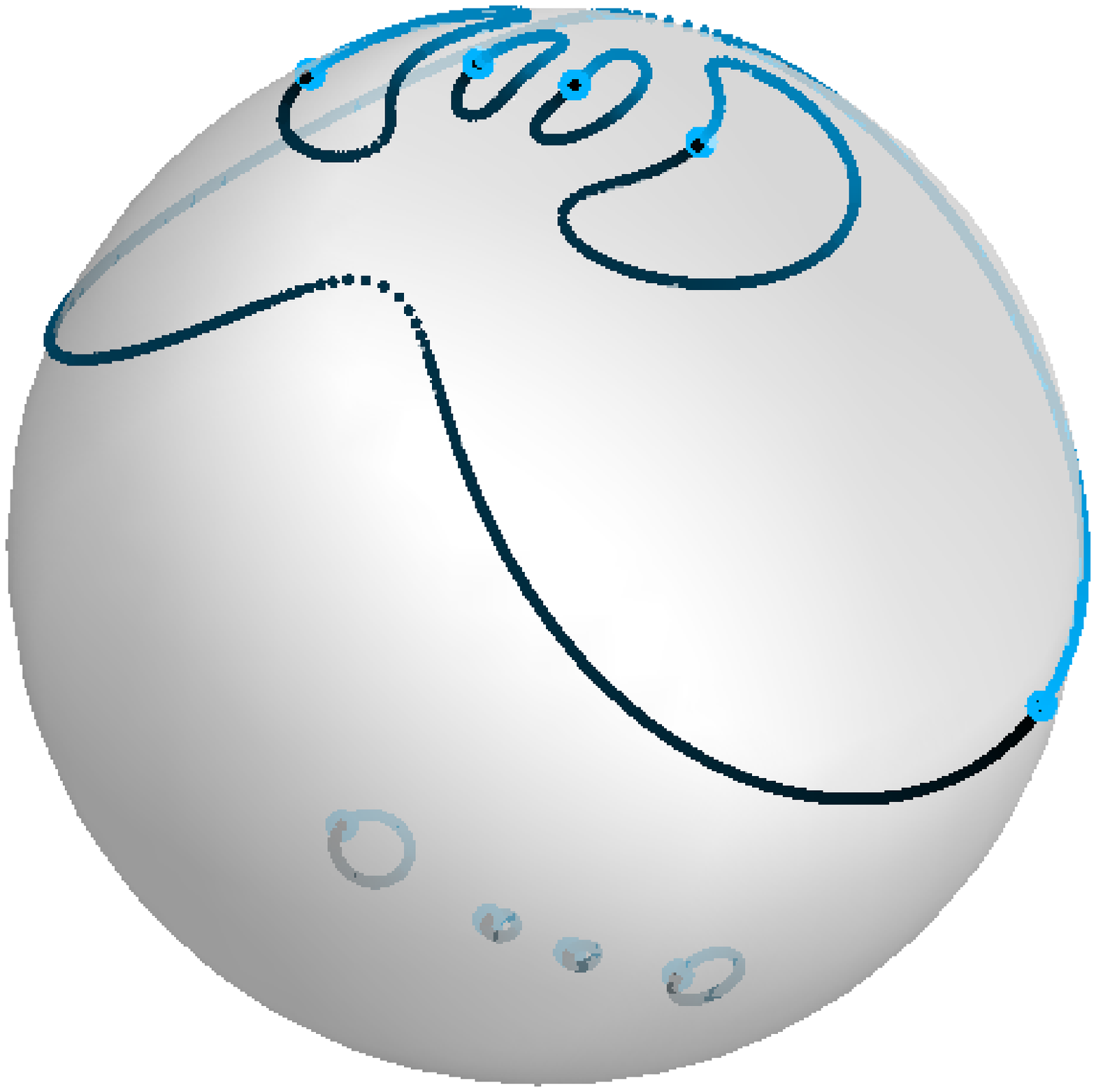}}
\hspace {15pt}
{\includegraphics[scale=0.37]{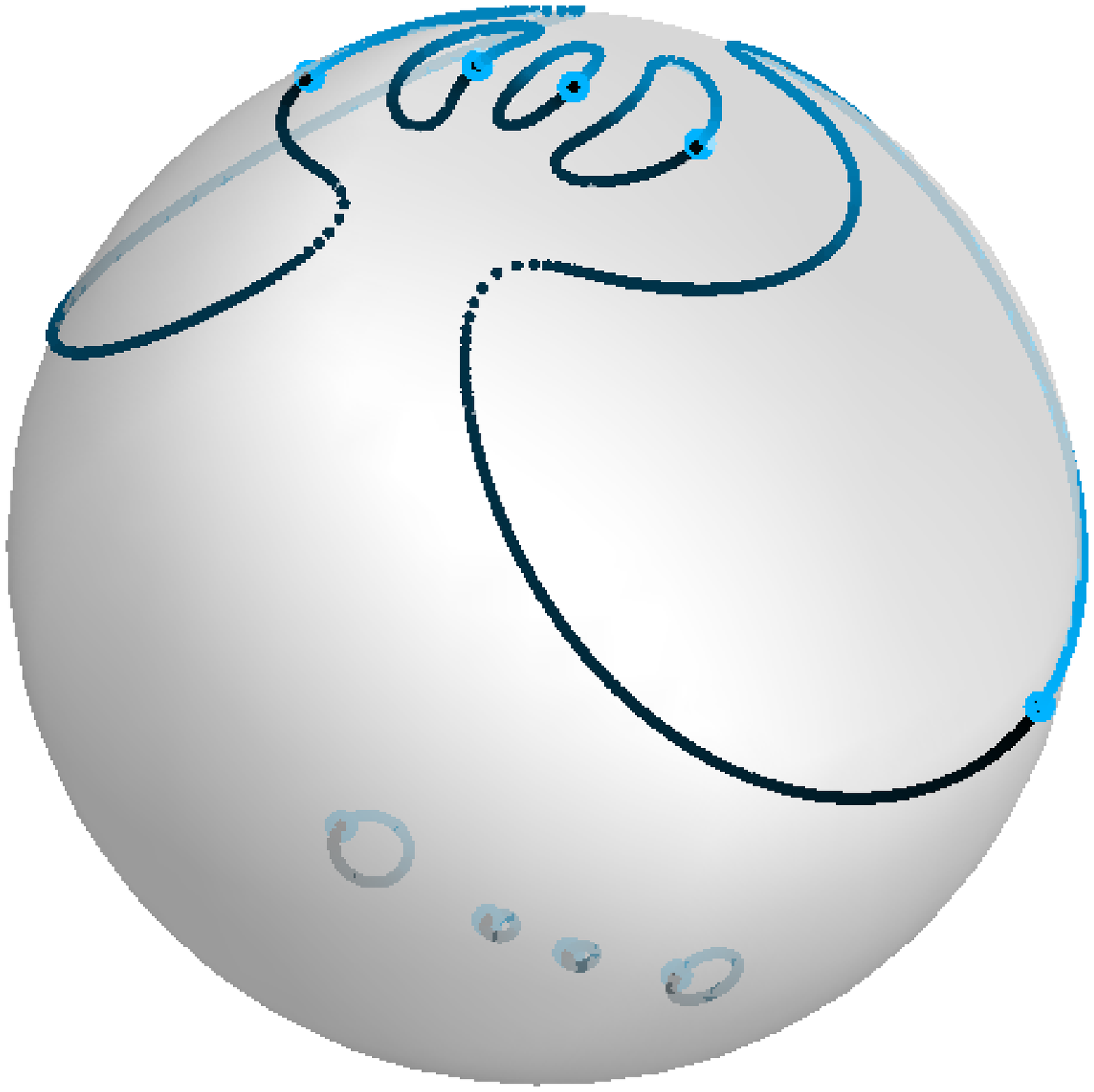}}
\end{center}
\caption{Trajectories of the zeros, in the corotating frame, for the highest energy level. Lighter blue corresponds to earlier in time, and darker to later. In the left figure the axis was chosen along $\eta=0.0345 \pi$, $\gamma=0$; and in the right figure the axis was along $\eta=0.0355 \pi$, $\gamma=0$. A small change in zenith angle has completely changed the topology of the trajectories.}
\label{FIG:zeros-xkett}
\end{figure}

The motion of the zeros  can be  quite complicated. The greatest intricacy  occurs for higher levels  where the Kramers pairs communicate more strongly. As an illustrative example  we plot in figure \ref{FIG:zeros-xkett} for one full cycle $\tau$, and for various axis orientations,  the trajectories in the corotating frame of the highest-level wavefunction zeros (polarized initially along $\hat{\bold{x}}$). Observe  that trajectories of nearby  axes can exhibit qualitatively different behavior --- in particular  closed cycles of individual zeros can combine into larger cycles in which the zeros permute locations. These trajectories have been calculated numerically by both using the projection method in Kato's equation, as well as by solving the equations of motion for the Majorana points, and are found to agree well.
(The intricate motion is best appreciated by viewing  animations.  We have made these available at \cite{movies})

Now  consider the time-dependence of the magnetization $\vev{\bold{J}}$.   By spinning the molecule at some frequency $1/T$ about the chosen axis, the magnetic moment may oscillate at a different frequency $1/ \tau$, which can be measured by placing pickup coils near the sample. The exact form of the moment oscillation depends on the energy level as well as the choice of initial state within the level.

\begin{figure}
\begin{center}
{\includegraphics[scale=0.5]{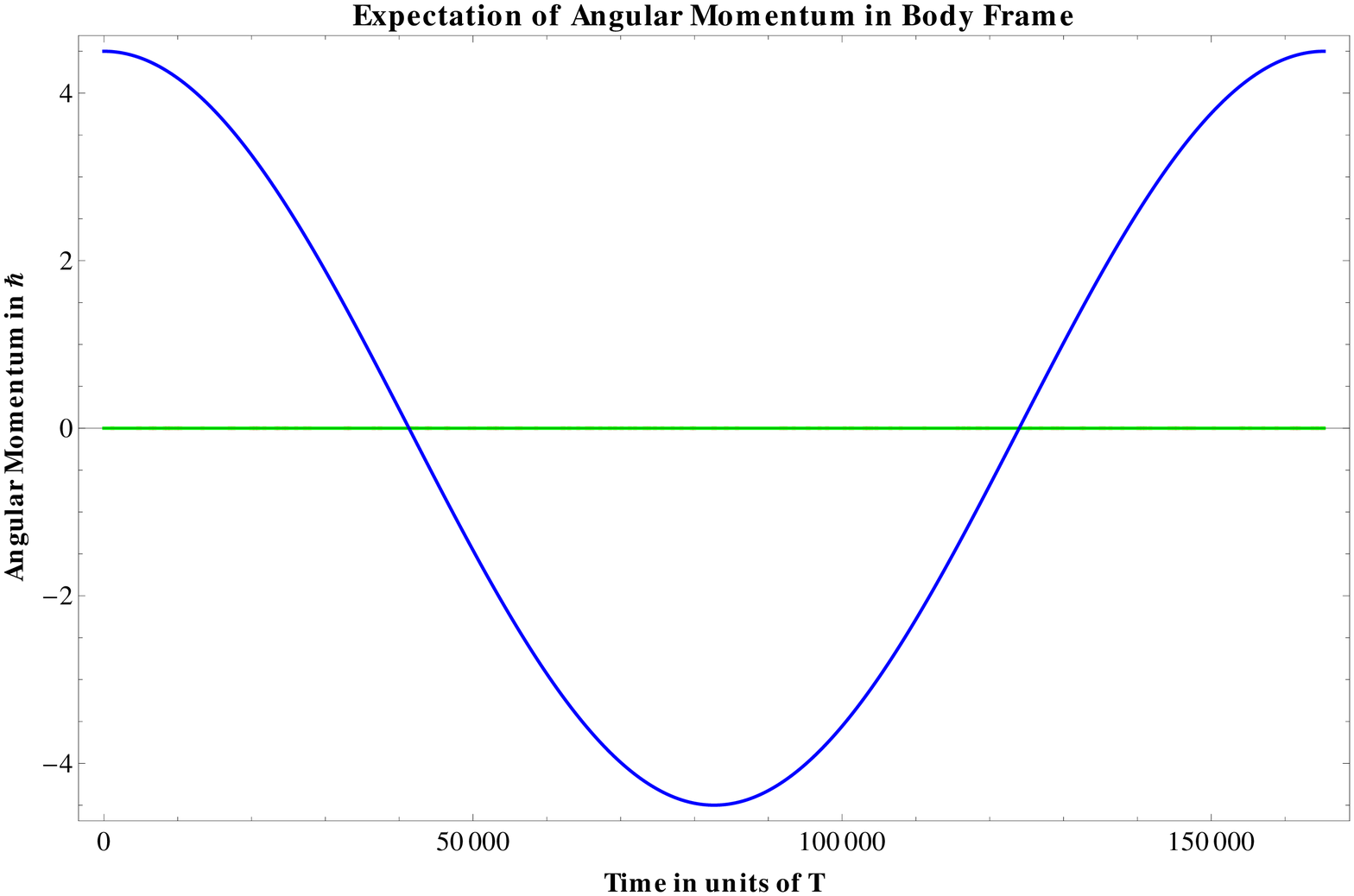}}
\end{center}
\caption{Plot of $\vev{\bold{J}(t)}$ for the ground state polarized along $\hat{\bold z}$, with $\vev{J_x}$ in red (not visible), $\vev{J_y}$ in green, and $\vev{J_z}$ in blue.}
\label{FIG:bodymu-zkett-axis1-gs}
\end{figure}

In figure \ref{FIG:bodymu-zkett-axis1-gs}  we look at a sample initially in the ground state, polarized along the $\hat{\bold{z}}$ direction. Such a state has a large initial magnetization along $\hat{\bold z}$ and zero magnetization along the transverse directions. We then rotate the sample about $\hat{\bold{x}}$ at a frequency of $2 \pi / T$, where $T$ is large compared the timescale of the gap --- in this case $\approx 8$ K, or $\approx 6 \times 10^{-12}$ seconds. We then calculate $\vev{\bold J}$ in the body frame of the molecule, and we find that while the $\vev{J_x}$ and $\vev{J_y}$ components do not vary much (being virtually zero), the $\vev{J_z}$ component exhibits appreciable oscillations on the timescale of $\tau \approx 165000 T$. What one should see in the lab frame then are rapid oscillations of $\vev{J_z}$ on the order of $T$, with an amplitude modulation  with  period  $\tau$. 

\begin{figure}
\begin{center}
{\includegraphics[scale=0.5]{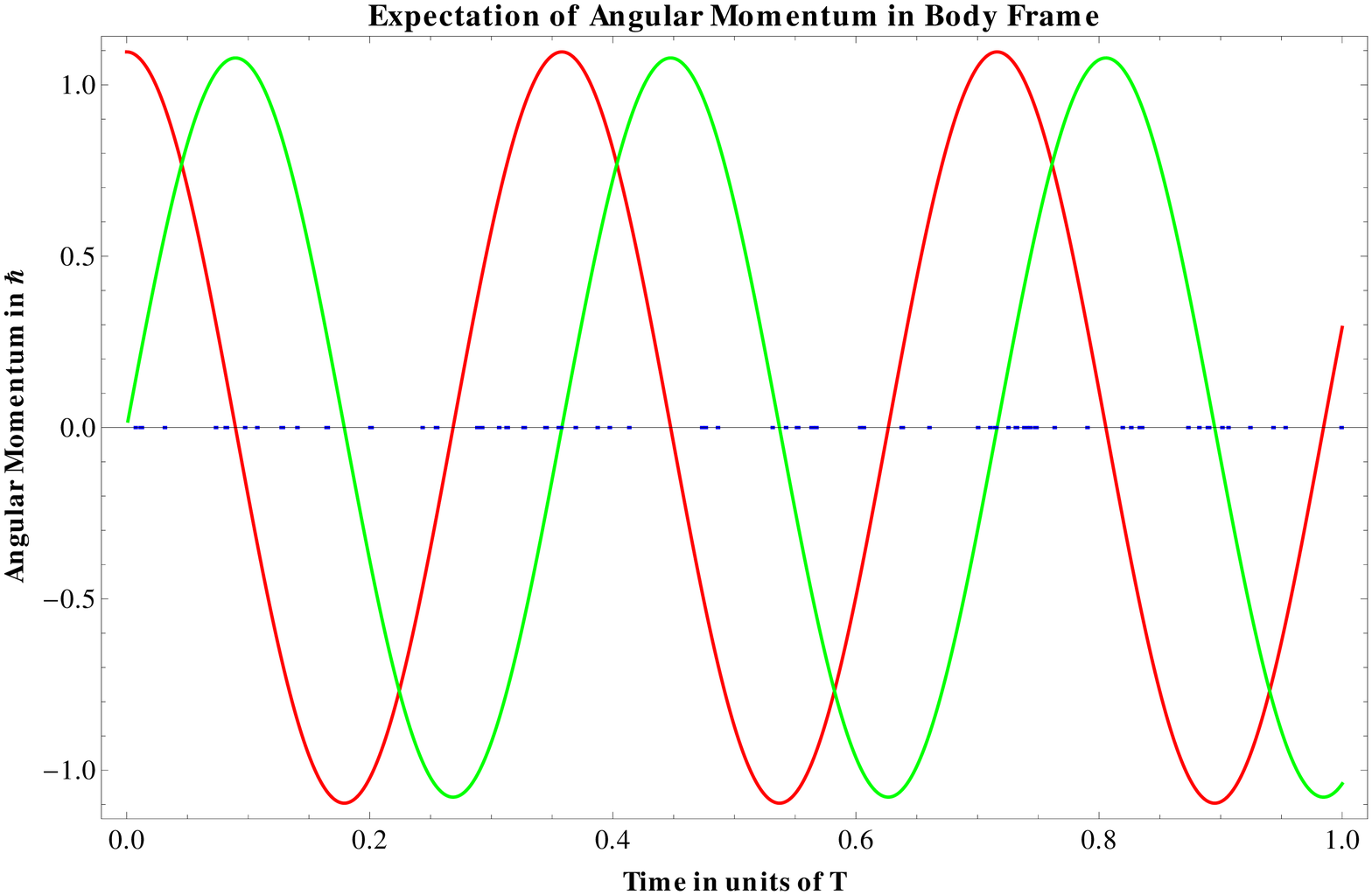}}\\
\vspace{12pt}
{\includegraphics[scale=0.5]{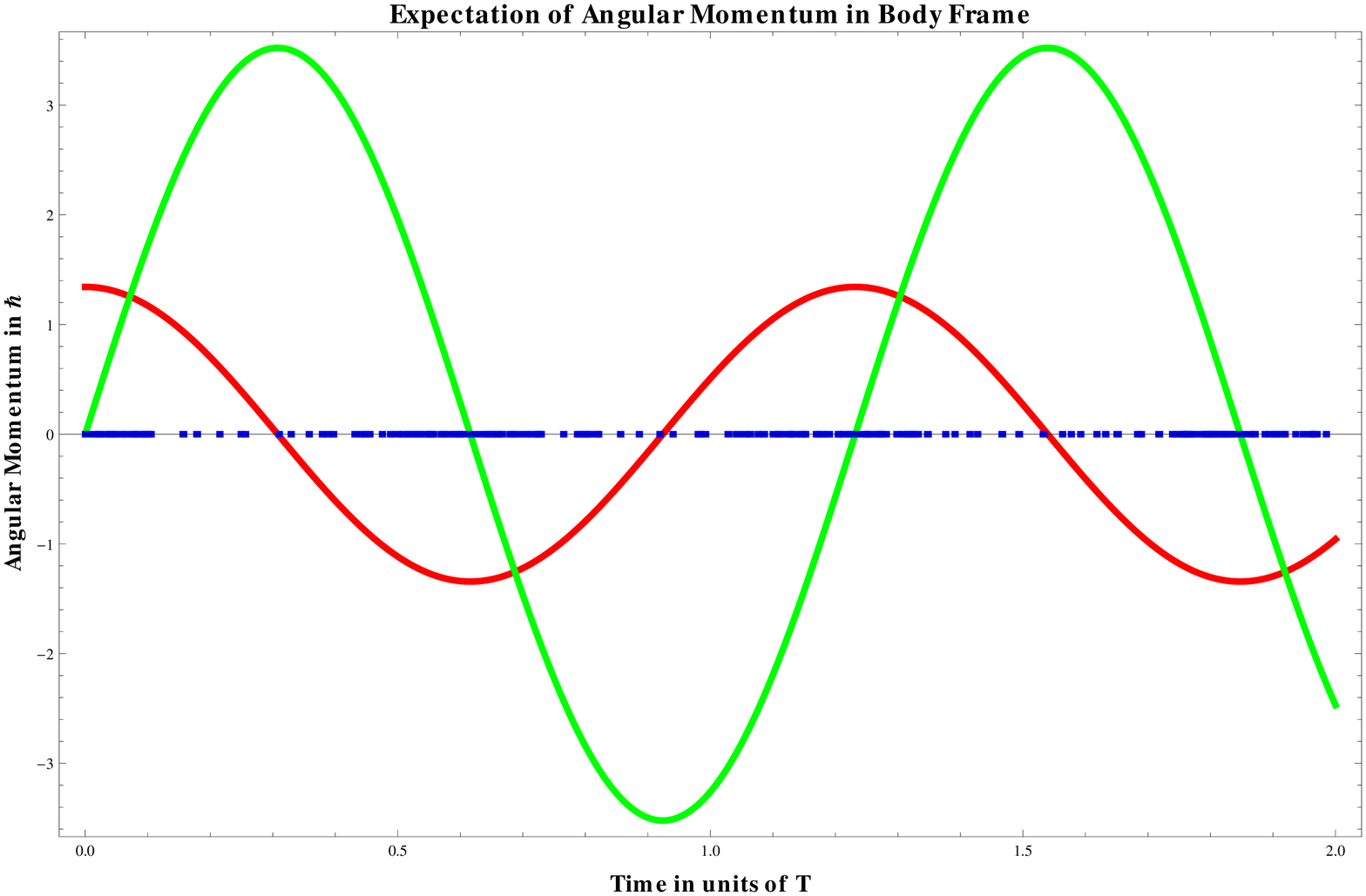}}
\end{center}
\caption{Plot of $\vev{\bold{J}(t)}$ for the third (top) and fourth (bottom) excited states polarized along $\hat{\bold x}$, with $\vev{J_x}$ in red, $\vev{J_y}$ in green, and $\vev{J_z}$ in blue.}
\label{FIG:bodymu-xkett-axis3-gs}
\end{figure}

Finally, in figure \ref{FIG:bodymu-xkett-axis3-gs} we look at a state  initially polarized along the $\hat{\bold{x}}$ direction and  rotated about $\hat{\bold z}$. The motivation for this is to look at states whose magnetization along the axis of rotation is suppressed. We find that as the  energy level becomes higher, the component $\vev{J_z}$ of the initial state becomes rapidly smaller , while $\vev{J_x}$ grows. We  expect that at higher levels non-Abelian effects  become stronger. Indeed we find that in the body frame, oscillations of the $\vev{J_x}$ and $\vev{J_y}$ components become more pronounced, which may be easier to detect. It would therefore be interesting if one can select an initial energy level, perhaps by applying a suitable a/c pulse to excite the molecule and then selecting the starting state within that level by applying a small d/c field which is gradually reduced to zero.

\section{discussion}

We have derived the equations of motions of   the Majorana points for states undergoing non-Abelian adiabatic evolution. We focussed primarily on the case when the degenerate subspace consists of a Karmers pair of eigenstates for a time-reversal invariant, half-integer spin system, and on  time-dependence that arises from simply rotating the host molecule about a  fixed axis.  We have seen that even when the effective  Hamiltonian is quite simple (quadratic anisotropy only) the resulting dance of the Majorana points can be quite intricate. When only one Majorana point is able to move its equation of motion is manifestly Hamiltonian. When many Majorana points are in motion the system is still Hamiltonian, but the associated symplectic form will be  complicated.

Non-Abelian evolution means that the motion of the period of the Hamiltonian (the rotation period)  and the period of the spin direction will in general be incommensurate.   The  resulting difference between the input  rotation frequency and the output magnetic moment frequency  can in principle be measured.  A confounding effect, however, is provided by the very moment that we are trying to measure. If it is too large, the resulting magnetic field will destroy  the degeneracy of the Kramers pair and will  wipe out the non-Abelian dynamics. If it is too small there will be no signal indicating the  spin direction. The splitting  of the degeneracy means that the two not-quite degenerate states will acquire a steadily  accumulating  relative phase that will ultimately overwhelm the non-Abelian affect.  To avoid this problem  we need to have a system  in which there is a substantial difference in scale between the small splitting of the near-degenerate states and the gap to the first excited state of the time-reversal invariant system.  We can then spin the system fast enough that the small splitting can be neglected, while the evolution within the Kramers pair eigenspace is still adiabatic. 

\section{Acknowledgements}

The work of YL was supported by grant NSF PHY08-55323AR, and that of MS  and AR by grant NSF DMR 09-03291.  We thank Anupam Garg and FeiFei Li for useful  discussions.

\end{document}